\documentclass[11pt,reqno]{extarticle}
\pdfoutput=1
\usepackage{geometry}
\usepackage{graphicx}

\usepackage{tocloft}
\usepackage{anyfontsize}
\usepackage[utf8]{inputenc}
\usepackage[english]{babel}

\usepackage{amssymb}
\usepackage{amsmath}
\usepackage{amsthm}

\usepackage{tikz}
\usetikzlibrary{arrows.meta}

\usepackage{pgfplots}

\usepackage{float}
\usepackage{caption}
\usepackage{subcaption}
\usepackage{scalerel}
\usepackage{enumitem}
\usepackage{hyperref}
\hypersetup{
  colorlinks   = true,
  urlcolor     = blue,
  linkcolor    = blue,
  citecolor   = red
}
\numberwithin{equation}{section}
\allowdisplaybreaks

\newtheorem{theorem}{Theorem}[section]
\newtheorem{definition}[theorem]{Definition}

\newtheorem{remark}[theorem]{Remark}
\newtheorem{lemma}[theorem]{Lemma}
\newtheorem{proposition}[theorem]{Proposition}
\newtheorem{example}[theorem]{Example}
\newtheorem{problem}[theorem]{Problem}

\newcommand{\ed}{\mathrm{d}}
\newcommand{\ringg}{\mathring{g}}
\newcommand{\dvol}{\mathrm{dvol}}

\title{\Large \textsc{Volume-Distance-Ratio Asymptote and Spacetime Inextendibility}}
\author{Pengyu Le}
\newcommand{\Address}{{
  \bigskip
  \footnotesize
  \textsc{Beijing Institute of Mathematical Sciences and Applications, Beijing, China}
  
  \textit{E-mail address}: \texttt{pengyu.le@bimsa.cn}
}}
\date{}

\begin{document}

\maketitle

\begin{abstract}
    This paper develops geometric criteria for determining the inextendibility of spacetimes near singularities based on asymptotic analysis of volume-distance relationships. We introduce and analyze the asymptotic behavior of the volume-distance-ratio (VDR), defined as the ratio of volumes of small chronological diamonds to appropriate powers of distances between their vertices. In $\mathrm{C}^0$ and $\mathrm{C}^{0,1}$ spacetimes (which are weaker than the classical $\mathrm{C}^2$ regularity), we prove that VDR converges to the Minkowski value as chronological diamonds approach accumulation points.

    The central contribution is the establishment of inextendibility criteria showing that failure of VDR convergence to the Minkowski value implies inextendibility of the spacetime. These criteria apply to spacetime extensions satisfying $\mathrm{C}^0$ locally null-non-accumulating strongly-causal conditions and $\mathrm{C}^{0,1}$ strongly-causal conditions, where the local null-non-accumulation condition is introduced as a fundamental structural property ensuring the validity of VDR-based inextendibility criteria.

    Concrete applications demonstrate the power and scope of these methods. We prove that $2$-dimensional Misner spacetime is $\mathrm{C}^0$ strongly-causal inextendible and that spatially flat FLRW spacetimes with linear scale factor behavior are $\mathrm{C}^0$ locally null-non-accumulating strongly-causal inextendible. Furthermore, we establish $\mathrm{C}^{0,1}$ strongly-causal inextendibility for Christodoulou's class of spherically symmetric self-similar naked singularity spacetimes.
\end{abstract}

\tableofcontents

%%%%%%%%%%%%%%%%%%%%%
%%%%%%%%%%%%%%%%%%%%%
%%%%%%%%%%%%%%%%%%%%%
\section{Introduction}

A fundamental problem in spacetime singularity theory concerns the manner in which spacetime terminates at the singularity. Since the Einstein equations are second-order covariant equations for the metric, classical $\mathrm{C}^2$ solutions break down when second-order metric invariants become discontinuous. When a spacetime fails to extend in a $\mathrm{C}^2$ manner, it may indicate inherent singularities with profound physical implications, such as curvature blow-up. A natural extension of this inquiry examines whether classical solutions admit extensions under weaker regularity conditions.

%%%%%%%%%%%%%%%%%%%%%
\subsection{VDR-based inextendibility criteria and application}
To study the extension problem under weaker regularity conditions, a natural approach is to examine geometric concepts that remain well-defined under such conditions, such as length and volume.

This paper introduces criteria for spacetime inextendibility by analyzing the asymptotic behavior of volumes and lengths near singularities. Moreover, it is natural to study geometric objects with canonical characterizations. We choose to investigate chronological diamonds near singularities. Roughly speaking, we use the volume-distance-ratio (VDR), a dimensionless quantity obtained from the ratio of the volume of small chronological diamonds to the power of the distance between their vertices, to detect inextendibility.

We investigate the VDR asymptote of small chronological diamonds under $\mathrm{C}^0$ and $\mathrm{C}^{0,1}$ regularities.

\begin{proposition}[See Proposition \ref{prop 5.5}]
    Let $(M,g)$ be a $(n+1)$-dimensional $\mathrm{C}^0$ globally hyperbolic spacetime and $p\in M$. Let $\{q_k\}\subset I^+(p)$ be a sequence converging to $p$ from a timelike direction. Define the volumes $V_k = \vert I(p,q_k) \vert$ and the distances $d_k = d(p,q_k)$.
    Then the {\bf VDR asymptote} of the chronological diamond $I(p,q_k)$ as $k\rightarrow +\infty$ is
    \begin{align*}
        \frac{V_k}{d_k^{n+1}}
        =
        \frac{2\omega_n}{n+1} \cdot \big( \frac{1}{2} \big)^{n+1} + o(1).
    \end{align*}
\end{proposition}

\begin{proposition}[See Proposition \ref{prop 5.6}]
    Let $(M,g)$ be a $(n+1)$-dimensional $\mathrm{C}^{0,1}$ globally hyperbolic spacetime and $p\in M$. Let $\gamma$ be a future-directed chronological geodesic emanating from $p$ which satisfies the geodesic equation
    $\ddot{\gamma}^{\kappa} + \Gamma_{\mu\nu}^{\kappa} \dot{\gamma}^{\mu} \dot{\gamma}^{\nu} = 0$,
    where $\Gamma_{\mu\nu}^{\kappa}$ and the differentials $\ed g_{\mu\nu}$ exist along $\gamma$. Define the volume $V(t) = \vert I(p,\gamma(t)) \vert$ and the length $l(t) = L[ \gamma|_{[0,t]} ]$.
    Then the {\bf VDR asymptote} of the chronological diamond $I(p,\gamma(t))$ as $t \rightarrow 0^+$ is
    \begin{align*}
        \frac{V(t)}{l^{n+1}(t)}
        =
        \frac{2\omega_n}{n+1} \cdot \big( \frac{1}{2} \big)^{n+1} + o(1).
    \end{align*}
\end{proposition}

We introduce the following rather general definition of local spacetime extension.

\begin{definition}[See Definitions \ref{def 2.4}, \ref{def 3.5}]
    Let $(M,g)$ be a globally hyperbolic spacetime and  $\hat{q}$ be a future boundary point. Let $\{ q_k \}$ be a future-ordering chronological sequence exhausting $\hat{q}$ in the chronological past. A {\bf local $\mathrm{C}^{k,\alpha}$ extension} of $(M,g)$ along $\{ q_k \}$ is a triple $(\tilde{M}, \tilde{g}, \phi)$ satisfying the following conditions:
    \begin{enumerate}[label=(\alph*)]
        \item
              A $\mathrm{C}^{k,\alpha}$ spacetime $(\tilde{M}, \tilde{g})$.

        \item
              $\phi: M \rightarrow \tilde{M}$, a smooth isometric embedding from $M$ to $\tilde{M}$.

        \item
              $Acc(\{\phi(q_k)\}) \neq \emptyset$: the set of accumulation points of $\{\phi(q_k)\}$ in $\tilde{M}$ is non-empty.
    \end{enumerate}
\end{definition}

The naive expectation is that the VDR asymptote of chronological diamonds approaching singularities could indicate inextendibility. However, examples demonstrate violations of this expected criterion.
\begin{example}[See Examples \ref{ex 4.22} \& \ref{ex 4.19}]
    \begin{enumerate}[label=\textbullet]
        \item
              $2$-dimensional Misner spacetimes $(U,\eta)$: $U = I^-(o) / \mathbb{Z}L$ where $L$ is a Lorentz boost and $\eta$ is the Minkowski metric (this example violates the expected criterion).
        \item
              $(V,g)$: $g=-(1-\lambda)^{\sigma} \ln(1-\lambda) \ed \sigma \ed \lambda$ and
              $V = \{ (\lambda,\sigma): \lambda\in (0,1), \sigma \in (2,3) \}$
              (this example illustrates the second obstruction situation below).
    \end{enumerate}
\end{example}
The following situations may occur when attempting to apply the VDR asymptote criterion:
\begin{enumerate}[label=\textbullet]
    \item
          Causality condition: The extension has less restrictive causality or continuity conditions.
    \item
          Failure of chronological-diamond-surjection (see Definition \ref{def 1.6}): The chronological future/past of the image of the singularities in the extension is not contained in the image of the spacetime. This obstruction invalidates the application of Proposition \ref{prop 5.5}, whose proof requires sequences converging from timelike directions.
\end{enumerate}
These obstructions are related to the local structure of the extension at the singularities. Motivated by the above analysis, we identify key structural properties of extensions that ensure the validity of the expected VDR asymptote criterion:
\begin{definition}[See Definitions \ref{def 2.3} \& \ref{def 4.12}]
    \begin{enumerate}[label=\textbullet]
        \item
              {\bf Strongly-causal condition}: There exists no arbitrary almost closed causal curve.
        \item
              {\bf Local null-non-accumulation condition}: Let $(M,g)$ be a $\mathrm{C}^2$ globally hyperbolic spacetime and $\hat{q}$ be a future boundary point. $(\tilde{M},\tilde{g},\phi)$ is a $\mathrm{C}^0$ strongly-causal extension of $(M,g)$ at $\hat{q}$. There exists a future-ordering chronological sequence $\{q_k\}$ exhausting $\hat{q}$ in the chronological past such that
              $\phi(\hat{q}) \notin Acc(\phi(E_+(q_k,M)))$,
              where $E^+(q_k,M)$ is the future horismos of $p_k$ in $(M,g)$.
    \end{enumerate}
\end{definition}

The causality condition is well-established in the literature. In contrast, to the author's knowledge, the local null-non-accumulation condition is introduced here for the first time in the study of spacetime extensions. This condition implies two useful properties concerning chronological relations in extensions: they ensure that the VDR asymptote at the singularities can be compared with the Minkowski value.
\begin{definition}[See Definitions \ref{def 4.13} \& \ref{def 4.18}]\label{def 1.6}
    Let $(M,g)$ be a $\mathrm{C}^2$ globally hyperbolic spacetime and $\hat{q}$ be a future boundary point. $(\tilde{M},\tilde{g},\phi)$ is a $\mathrm{C}^0$ strongly-causal extension of $(M,g)$ at $\hat{q}$. Denote $\phi(\hat{q})$ by $\tilde{q}$.
    \begin{enumerate}[label=\textbullet]
        \item
              {\bf Local chronological-diamond-preservation} at $\hat{q}$: there exists a neighbourhood $V$ of $\hat{q}$, such that for any $p,q\in V$,
              $I(\phi(p), \phi(q), \tilde{M})
                  =
                  \phi(I(p,q,V))$.
        \item
              {\bf Local past-chronological-diamond-surjection}: there exist future-ordering chronological sequences
              $\{q_k\}$ exhausting $\hat{q}$ in the chronological past,
              and $\{\tilde{p}_k\} \subset I^-(\tilde{q},\tilde{M})$ converging to $\tilde{q}$,
              such that
              $I(\tilde{p}_k,\tilde{q},\tilde{M}) \subset \phi(I(q_k,\hat{q}))$.
    \end{enumerate}
\end{definition}

We then obtain the following criterion for $\mathrm{C}^0$ spacetime extension using the VDR asymptote.

\begin{theorem}[{\bf $\mathrm{C}^0$ locally null-non-accumulating strongly-causal inextendibility}, see Theorem \ref{thm 6.1}]
    Let $(M,g)$ be a $\mathrm{C}^2$ globally hyperbolic spacetime and $\hat{q}$ be a future boundary point. There exists no local $\mathrm{C}^0$ locally null-non-accumulating strongly-causal extension of $I^-(\hat{q})$ at $\hat{q}$ if for any future-ordering chronological sequence $\{ q_k \}$ exhausting $\hat{q}$ in the chronological past, we have that
    \begin{align*}
        \frac{V_k}{ d_k^{n+1} } \nrightarrow \frac{2\omega_n}{n+1} \cdot (\frac{1}{2})^{n+1},
    \end{align*}
    where
    $V_k = \vert I(q_k, \hat{q}) \vert$,
    $d_k = d(q_k, \hat{q})$.
\end{theorem}

Notably, for $\mathrm{C}^{0,1}$ extensions, the local null-non-accumulation condition is automatically satisfied (see Proposition \ref{prop 4.26}). This yields the following rigorous criterion for $\mathrm{C}^{0,1}$ spacetime extensions.

\begin{theorem}[{\bf $\mathrm{C}^{0,1}$ strongly-causal inextendibility}, see Theorem \ref{thm 6.2}]
    Let $(M,g)$ be a $\mathrm{C}^2$ globally hyperbolic spacetime and $\hat{q}$ be a future boundary point. There exists no local $\mathrm{C}^{0,1}$ strongly-causal extension of $I^-(\hat{q})$ at $\hat{q}$ if there exists a future-directed chronological geodesic curve $\gamma: [t_0,0) \rightarrow M$ exhausting $\hat{q}$ in the chronological past, such that as $t\rightarrow 0^-$,
    \begin{align*}
        \frac{V(t)}{ l(t)^{n+1} } \nrightarrow \frac{2\omega_n}{n+1} \cdot (\frac{1}{2})^{n+1},
    \end{align*}
    where
    $V(t) = \vert I(\gamma(t), \hat{q}) \vert$,
    $l(t) = L[\gamma|_{[t,0)}]$.
\end{theorem}

To demonstrate the applicability of these criteria, we apply them to study spacetime extensions of several specific spacetimes.

\begin{theorem}[See Theorem \ref{thm 6.3}]
    The $2$-dimensional Misner spacetime admits no $\mathrm{C}^0$ strongly-causal extension at the future boundary point $o$.
\end{theorem}
Note that while the $2$-dimensional Misner spacetime admits smooth spacetime extensions containing closed causal curves (thus violating strongly causal condition). This result highlights the crucial role of the strongly causal condition in spacetime extensions.

The volume-distance-ratio asymptote criteria establish the $\mathrm{C}^0$ inextendibility of the following spatially flat FLRW spacetimes at the future timelike boundary. Reversing the time orientation implies $\mathrm{C}^0$ inextendibility for spatially flat FLRW spacetimes with big bang singularities.
\begin{theorem}[See Theorem \ref{thm 7.1}]
    The spatially flat FLRW spacetime $(M = \mathbb{R}^{n+1}_{t<0},g)$ with $n\geq 2$ and
    \begin{align*}
        g
        =
        - \ed t^2 + a^2(t) ( \ed r^2 + r^2 \ed \sigma^2 ),
        \quad
        a(t) \sim |t|.
    \end{align*}
    admits no local $\mathrm{C}^0$ locally null-non-accumulating strongly-causal extension at the future timelike boundary point $\mathcal{O}$.
\end{theorem}

The volume-distance-ratio asymptote criteria also establish the $\mathrm{C}^{0,1}$ inextendibility of the following class of naked singularity spacetimes.
\begin{theorem}[See Theorem \ref{thm 8.5}]
    For $0< k <1$, the interior solution of the spherically self-similar naked singularity spacetime constructed by Christodoulou in \cite{Chr94} admits no $\mathrm{C}^{0,1}$ strongly-causal extension at the scaling origin.
\end{theorem}

%%%%%%%%%%%%%%%%%%%%%
\subsection{Previous results}
The spacetime inextendibility problem is intimately connected with the strong  cosmic censorship conjectures regarding singularities in general relativity, proposed by Penrose in \cite{P79}. Such singularities may compromise the predictability of observations by local observers. The conjecture was therefore aimed at preventing the formation of such singularities. A mathematical formulation states that generic asymptotically flat initial data yield a maximal future development locally inextendible in suitable weak regularity, for instance, continuously in \cite{Chr99b}, or with locally $L^2$-integrable Christoffel symbols in \cite{Chr09} following Christodoulou. From a PDE perspective, inextendibility with locally $L^2$-integrable Christoffel symbols implies that no extension exists as a solution—even in a weak sense—that includes any portion of the boundary.

In a series of papers \cite{Chr91}-\cite{Chr99a}, Christodoulou proved a version of the strong cosmic censorship conjecture for the spherically symmetric Einstein-scalar field system. He demonstrated that generically, the maximal globally hyperbolic development is $\mathrm{C}^0$ inextendible within spherically symmetric Lorentzian manifolds. He also established the weak cosmic censorship for this system. An alternative approach to the weak cosmic censorship in this system via interior perturbations was introduced by Li in \cite{Li25}.

A further breakthrough was Christodoulou's proof of black hole formation in vacuum without symmetry assumptions \cite{Chr09}, which generalized his earlier results \cite{Chr91}. This motivated investigations of weak cosmic censorship in gravitational collapse without symmetries by Li and Liu in \cite{LL18} \cite{LL22}, and by An in \cite{A25}.

The first demonstration of $\mathrm{C}^0$ inextendibility without symmetry assumptions was Sbierski's result for the Schwarzschild solution in \cite{Sb18a} \cite{Sb18b}. In \cite{GL17}, Galloway and Ling established the $\mathrm{C}^0$ inextendibility of AdS spacetime and showed $\mathrm{C}^0$ extendibility for a class of hyperbolic ($K=-1$) FLRW spacetimes, termed Milne-like. In \cite{GLS18}, Galloway, Ling and Sbierski proved that global hyperbolicity with timelike geodesic completeness implies $\mathrm{C}^0$ inextendibility. These results were extended to various settings in \cite{GL18} (timelike geodesically complete spacetimes), \cite{MS19} (Lorentz–Finsler spaces), \cite{GKS19} (Lorentzian length spaces), \cite{Lin20} ($K=-1$ FLRW spacetimes), \cite{Sb23} ($K=\pm 1$ FLRW spacetimes), \cite{Lin26} ($K=0$ FLRW spacetimes), \cite{Mie24} (Kasner spacetimes).

For other models, in \cite{D03}, Dafermos showed that solutions to the spherically symmetric Einstein-Maxwell-scalar field system with initial data near the Reissner-Nordström solution yield $\mathrm{C}^0$ extendible maximal globally hyperbolic developments. He established that generically, the Hawking mass blows up, implying inextendibility within spherically symmetric Lorentzian manifolds having locally $L^2$-integrable Christoffel symbols. In \cite{LO19a} \cite{LO19b}, Luk and Oh proved the $\mathrm{C}^2$ strong cosmic censorship for the Einstein-Maxwell-scalar-field system in spherical symmetry with two-ended asymptotically flat data, demonstrating generic inextendibility in $\mathrm{C}^2$ metrics. In \cite{Sb22}, Sbierski established $\mathrm{C}_{\text{loc}}^{0,1}$ inextendibility of spherically symmetric weak null singularities via holonomy computations.

In \cite{Lu18}, Luk constructed weak null singularities with non-locally $L^2$ Christoffel symbols. In \cite{DL25}, Dafermos and Luk proved $\mathrm{C}^0$ stability of the Kerr-Cauchy horizon without symmetry assumptions and generic $L^2$ blow-up of the Christoffel symbols. In \cite{Sb25}, Sbierski demonstrated $\mathrm{C}_{\text{loc}}^{0,1}$ inextendibility for weak null singularities without symmetry assumptions, introducing a strategy to infer $\mathrm{C}_{\text{loc}}^{0,1}$ inextendibility from curvature blow-up. Recently, \cite{Gu26} and \cite{LS26} further investigated the formation of weak null singularities and the (in)stability of the Kerr–Cauchy horizon for generic rotating black holes, applying the strategy from \cite{Sb25} to establish $\mathrm{C}_{\text{loc}}^{0,1}$ inextendibility at the Kerr–Cauchy horizon. Weak null singularities have also been constructed for the Einstein–Euler system by Song in \cite{So25}.

Spacetime extensions are closely tied to the initial value problem for the Einstein equations. Local well-posedness was first established by Choquet-Bruhat in \cite{Cho52}. Significant efforts addressed lowering the regularity requirements. In \cite{KR10} \cite{Wq12}, Klainerman, Rodnianski and Wang derived breakdown criteria for rough solutions of the Einstein equations. Pursuing critical regularity culminated in the resolution of the bounded $L^2$ curvature conjecture by Klainerman, Rodnianski and Szeftel in \cite{KRS15}.

Volumes of small causal diamonds in smooth settings have been studied primarily in the physics literature, such as in \cite{My78}, \cite{GS07}, \cite{J17}, \cite{Wj19}. In \cite{Le23}, the author demonstrated that causal diamonds solve a type of isoperimetric problem for the domain of dependence in Minkowski spacetime. Numerous studies exist on Lorentzian geometry in low regularity, including causality theory, global hyperbolicity, geodesic theory, and singularity theorems. \cite{CG12} demonstrated bubbling sets in $\mathrm{C}^0$ spacetimes to illuminate differences between continuous and smooth Lorentzian causality theory, a result closely related to Example \ref{ex 4.19}. \cite{Min19} generalized causality theory under weak assumptions. \cite{CG12} \cite{FS12} \cite{Sa16} investigated global hyperbolicity in low regularity from various perspectives. \cite{GL17} proved that maximal curves are either lightlike or timelike almost everywhere in $\mathrm{C}^{0,1}$ Lorentzian manifolds. \cite{LLS21} strengthened this by showing maximal causal curves in $\mathrm{C}^{0,1}$ Lorentzian manifolds admit $\mathrm{C}^{1,1}$-parameterizations and solved the geodesic equation in the sense of Filippov. Classical singularity theorems have been generalized to $\mathrm{C}^{1,1}$ and $\mathrm{C}^1$ regularities in \cite{KSV15} \cite{KSSV15} \cite{GGKS18} \cite{Gr20} \cite{KOSS22}.

%%%%%%%%%%%%%%%%%%%%%
%%%%%%%%%%%%%%%%%%%%%
%%%%%%%%%%%%%%%%%%%%%
\section{Causal structure and spacetime extension}

%%%%%%%%%%%%%%%%%%%%%
%%%%%%%%%%%%%%%%%%%%%
\subsection{Causal structure}
\begin{definition}\label{def 2.1}
    A $\mathrm{C}^{k,\alpha}$ spacetime $(M,g)$ is a time-oriented $\mathrm{C}^{k,\alpha}$ Lorentzian manifold (a smooth manifold $M$ equipped with a $\mathrm{C}^{k,\alpha}$ Lorentzian metric $g$).
\end{definition}

The causal structure is a fundamental concept in spacetime theory. Following the approach of \cite{GKSS20}, we adopt the following definitions for the causality theory.
\begin{definition}\label{def 2.2}
    Let $(M,g)$ be a spacetime.
    \begin{enumerate}[label=\textbullet]
        \item
              $p\ll q$: $p$ chronologically precedes $q$ if there exists a smooth future-directed timelike curve from $p$ to $q$.
        \item
              $p < q$: $p$ strictly causally precedes $q$ if there exists a locally Lipschitz future-directed causal curve from $p$ to $q$.
        \item
              $p \prec q$: $p$ causally precedes $q$ if $p < q$ or $p=q$.
        \item
              $\ll_U, <_U, \prec_U$: chronological, strict causal, and causal precedence relations in $U$.
        \item
              $I^{\pm}(p)$: chronological future/past of $p$. $I^{\pm}(p) = \{ q\in M: p \ll q/q\ll p\}$.
        \item
              $J^{\pm}(p)$: causal future/past of $p$. $J^{\pm}(p) = \{ q\in M: p \prec q / q \prec q\}$.
        \item
              $I^{\pm}(V)$: chronological future/past of $V\subset M$. $I^{\pm}(V) = \cup_{p\in V} I^{\pm}(p)$.
        \item
              $J^{\pm}(V)$: causal future/past of $V\subset M$. $J^{\pm}(V) = \cup_{p\in V} J^{\pm}(p)$.
        \item
              $\mathcal{D}^{\pm}(V)$: future/past domain of dependence of $V$,
              \begin{align*}
                  \mathcal{D}^{\pm}(V)
                  =
                  \{ q:
                   &
                  \text{every past/future-inextendible causal curve }
                  \\
                   &
                  \text{through $q$ interests $V$} \}
              \end{align*}
        \item
              $I^{\pm}(p,U), J^{\pm}(p,U)$: chronological future/past and causal future/past of $p$ in $(U,g)$ where $p\in U\subset M$.
        \item
              $I^{\pm}(V,U)$: chronological future/past of $V$ in $(U,g)$. $I^{\pm}(V,U) = \cup_{p\in V\cap U} I^{\pm}(p,U)$.
        \item
              $J^{\pm}(V,U)$: causal future/past of $V$ in $(U,g)$. $J^{\pm}(V,U) = \cup_{p\in V\cap U} J^{\pm}(p,U)$.
        \item
              $\mathcal{D}^{\pm}(V,U)$: future/past domain of dependence of $V\cap U$ in $U$.
        \item
              $I(p,q)$: chronological diamonds between $p,q$. $I(p,q) = I^+(p) \cap I^-(q)$.
        \item
              $J(p,q)$: causal diamonds between $p,q$. $J(p,q) = J^+(p) \cap J^-(q)$.
        \item
              $I(p,q,U)$, $J(p,q,U)$: chronological and causal diamonds between $p,q$ in $(U,g)$ where $U\subset M$.
    \end{enumerate}
\end{definition}

%%%%%%%%%%%%%%%%%%%%
%%%%%%%%%%%%%%%%%%%%%
\subsection{Causality condition and spacetime extension}
We introduce several types of causality conditions for spacetimes.
\begin{definition}\label{def 2.3}
    Let $(M,g)$ be a spacetime.
    \begin{enumerate}[label=\textbullet]
        \item
              Global hyperbolicity: There exists a Cauchy hypersurface.

        \item
              Strongly causal spacetime: For any neighbourhood $U$ of $p\in M$, there exists a neighbourhood $V\subset U$, $p\in V$ such that $V$ is causally convex in $M$.

        \item
              Future/Past-distinguishing spacetime: $I^{\pm}(p) = I^{\pm}(q) \Rightarrow p=q$.

    \end{enumerate}
\end{definition}
We refer to \cite{MS08} for detailed discussions on causality conditions and the causal hierarchy in $\mathrm{C}^2$ spacetimes. For global hyperbolicity in $\mathrm{C}^0$ spacetimes, see \cite{Sa16}.

We now introduce the notion of local $\mathrm{C}^{k,\alpha}$ spacetime extensions with various causality conditions.
\begin{definition}[{\bf Local $\mathrm{C}^{k,\alpha}$ $\mathcal{C}$(causality condition) extension} along a divergent sequence escaping any compact set]\label{def 2.4}
    Let $(M,g)$ be a globally hyperbolic spacetime.
    $\{ q_k \}$ is a divergent sequence escaping any compact set in $M$, i.e. for any compact set $K\subset M$, there exists $N_K$ such that $q_{k} \notin K$ for $k \geq N_K$.
    A local $\mathrm{C}^{k,\alpha}$ extension of $(M,g)$ along $\{ q_k \}$ is a triple $(\tilde{M}, \tilde{g}, \phi)$ satisfying the following conditions:
    \begin{enumerate}[label=(\alph*)]
        \item
              A $\mathrm{C}^{k,\alpha}$ spacetime $(\tilde{M}, \tilde{g})$.

        \item
              $\phi: M \rightarrow \tilde{M}$, a smooth isometric embedding from $M$ to $\tilde{M}$.

        \item
              $Acc(\{\phi(q_k)\}) \neq \emptyset$: the set of accumulation points of $\{\phi(q_k)\}$ in $\tilde{M}$ is non-empty.
    \end{enumerate}
    If $(\tilde{M},\tilde{g})$ additionally satisfies the causality condition $\mathcal{C}$, then $(\tilde{M}, \tilde{g}, \phi)$ is called a local $\mathrm{C}^{k,\alpha}$ $\mathcal{C}$ extension of $(M,g)$ along $\{ q_k \}$.
\end{definition}

\begin{definition}[{\bf Local $\mathrm{C}^{k,\alpha}$ ($\mathcal{C}$) inextendibility}]\label{def 2.5}
    Let $(M,g)$ be a globally hyperbolic spacetime and $\{ q_k \}$ be a divergent sequence escaping any compact set.
    $(M,g)$ is locally $\mathrm{C}^{k,\alpha}$ ($\mathcal{C}$) inextendible along $\{q_k\}$ if $(M,g)$ admits no local $\mathrm{C}^{k,\alpha}$ ($\mathcal{C}$) extension along $\{ q_k \}$.
    $(M,g)$ is locally $\mathrm{C}^{k,\alpha}$ ($\mathcal{C}$) inextendible if it is locally $\mathrm{C}^{k,\alpha}$ ($\mathcal{C}$) inextendible along every divergent sequence that escapes every compact set.
\end{definition}

%%%%%%%%%%%%%%%%%%%%%
%%%%%%%%%%%%%%%%%%%%%
%%%%%%%%%%%%%%%%%%%%%
\section{Strongly-causal extension}

%%%%%%%%%%%%%%%%%%%%%
%%%%%%%%%%%%%%%%%%%%%
\subsection{Future boundary}
We introduce a construction of the spacetime boundary based on causal structure, which essentially corresponds to the construction of ideal points via \textit{terminal indecomposable future/past-sets (TIFs/TIPs)} as introduced by \cite{GKP72}.

\begin{definition}[{\bf Future boundary}]\label{def 3.1}
    Let $(M,g)$ be a spacetime.
    \begin{enumerate}[label=\textbullet]
        \item
              $\mathcal{FS}  = \{ \{q_k\}: q_k \ll q_{k+1} \}$:
              the set of future-ordering chronological sequences.

        \item
              $\mathcal{FS}_{\sim}$: the set of equivalence classes in $\mathcal{FS}$ where the equivalence relation $\sim$ on $\mathcal{FS}$is defined by
              $\{q_k\} \sim \{ q'_{k} \}
                  \Leftrightarrow
                  I^-(\{q_k\}) = I^-(\{q'_{k}\})$.
              Denote by $[q_k]$ the equivalence class of $\{q_k\}$.

        \item
              $I^-(\hat{q})$, $\hat{q} \in \mathcal{FS}_{\sim}$: chronological past of $\hat{q}$,
              $I^-(\hat{q}) = I^-(\{q_k\})$ where $[q_k] = \hat{q}$.

        \item
              A future-ordering chronological sequence $\{q_k\}$ in $(M,g)$ is said to exhaust $\hat{q}$ in the chronological past if $\hat{q} = [q_k]$.\footnote{For brevity, we omit the phrase ``in the chronological past" when the context is clear.}

        \item
              $I^+(\hat{q}, \mathcal{FS}_{\sim})=
                  \{ \hat{p} \in \mathcal{FS}_{\sim} :
                  \exists p \in I^-(\hat{p}),
                  \text{ s.t. }
                  I^-(\hat{q}) \subset I^- (p)
                  \}$: chronological future of $\hat{q}$ in $\mathcal{FS_{\sim}}$.
        \item
              $\mathcal{FB}
                  =
                  \{ \hat{q} \in \mathcal{FS}_{\sim}: I^+(\hat{q},\mathcal{FS_{\sim}}) = \emptyset \}$:
              future boundary of $(M,g)$ defined as the subset of $\mathcal{FS}_{/\sim}$ consisting of elements with empty chronological future.
    \end{enumerate}
\end{definition}

\begin{definition}[{\bf Chronological diamond ${I(p,\hat{q})}$}]\label{def 3.2}
    Let $\hat{q}$ be a future boundary point of $(M,g)$. Suppose $p \in I^-(\hat{q})$. Define the chronological diamond between $p$ and $\hat{q}$, denoted by $I(p, \hat{q})$, as
    $I(p,\hat{q})
        =
        I^+(p) \cap I^-(\hat{q})$.
\end{definition}

\begin{definition}\label{def 3.3}
    Let $\hat{q}$ be a future boundary point of $(M,g)$.
    \begin{enumerate}[label=\textbullet]
        \item
              An open subset $U$ of $I^-(\hat{q})$ is called a neighbourhood of $\hat{q}$ in $I^-(\hat{q})$ if there exist a representative $\{q_k\} \in \hat{q}$ and $k$ such that $I(q_k,\hat{q}) \subset U$.
        \item
              Let $\gamma$ be a future-directed timelike curve in $(M,g)$. We say that $\gamma$ exhausts $\hat{q}$ in the chronological past if $I^-(\gamma) = I^-(\hat{q})$.
    \end{enumerate}
\end{definition}

\begin{definition}[{\bf Distance $d(p,\hat{q})$}]\label{def 3.4}
    Let $\hat{q}$ be a future boundary point of $(M,g)$. Suppose $p \in I^-(\hat{q})$. Define the distance between $p$ and $\hat{q}$, denoted by $d(p, \hat{q})$, as
    $d(p,\hat{q})
        =
        \sup\{ \lim_{k \rightarrow +\infty} d(p,q_k): \text{$\{q_k\}$ exhausts $\hat{q}$ in the chronological past} \}$.
\end{definition}

\begin{definition}[{\bf Local $\mathrm{C}^{k,\alpha}$ ($\mathcal{C}$) inextendibility at $\hat{q}$}]\label{def 3.5}
    Let $(M,g)$ be a globally hyperbolic spacetime with a future boundary point $\hat{q}$.
    $(M,g)$ is locally $\mathrm{C}^{k,\alpha}$ ($\mathcal{C}$) inextendible at $\hat{q}$ if $(M,g)$ admits no local $\mathrm{C}^{k,\alpha}$ ($\mathcal{C}$) extension along any future-ordering chronological sequence $\{q_k \}$ exhausting $\hat{q}$.
\end{definition}

%%%%%%%%%%%%%%%%%%%%%
%%%%%%%%%%%%%%%%%%%%%
\subsection{Future-boundary-continuous (FBC) extension}

\begin{definition}[{\bf Local $\mathrm{C}^{k,\alpha}$ FBC extension} of {$I^-(\hat{q})$} at {$\hat{q}$}]\label{def 3.6}
    Let $(M,g)$ be a globally hyperbolic spacetime with a future boundary point $\hat{q}$. A local $\mathrm{C}^{k,\alpha}$ FBC extension of $I^-(\hat{q})$ at $\hat{q}$ is a triple $(\tilde{M}, \tilde{g}, \phi)$ satisfying:
    \begin{enumerate}[label=(\alph*)]
        \item $(\tilde{M}, \tilde{g})$ is a $\mathrm{C}^{k,\alpha}$ globally hyperbolic spacetime.

        \item $\phi$ is a smooth isometric embedding from a neighbourhood $U$ of $\hat{q}$ in $I^-(\hat{q})$ to $\tilde{M}$.

        \item
              FBC at $\hat{q}$: $\phi$ extends continuously to $\hat{q}$, i.e. $\lim_{k\rightarrow +\infty} \phi(q_k)$ exists, denoted by $\phi(\hat{q})$, for any future-ordering chronological sequence $\{q_k\}$ exhausting $\hat{q}$.
    \end{enumerate}
    If there exists (no) local $\mathrm{C}^{k,\alpha}$ FBC extension of $I^-(\hat{q})$ at $\hat{q}$, we say that $I^-(\hat{q})$ is locally $\mathrm{C}^{k,\alpha}$ FBC (in)extendible at {$\hat{q}$}.
\end{definition}

\begin{definition}[{\bf Local $\mathrm{C}^{k,\alpha}$ ($\mathcal{C}$-)FBC extension} at $\hat{q}$]\label{def 3.7}
    Let $(M,g)$ be a globally hyperbolic spacetime with a future boundary point $\hat{q}$.
    Let $(\tilde{M},\tilde{g},\phi)$ be a local $\mathrm{C}^{k,\alpha}$ ($\mathcal{C}$) extension of $(M,g)$ along a representative $\{q_k\}$ of $\hat{q}$. $\phi$ is FBC at $\hat{q}$ if it satisfies condition (c) of Definition \ref{def 3.6}.
\end{definition}

\begin{remark}\label{rem 3.8}
    Local $\mathrm{C}^{k,\alpha}$ FBC extendibility of $(M,g)$ at $\hat{q}$ implies local $\mathrm{C}^{k,\alpha}$ FBC extendibility of $I(\hat{q})$ at $\hat{q}$ (see Proposition \ref{prop 3.9} below).

    Recall that the past-distinguishing causality condition states that $I^-(p) = I^-(q)$ implies $p=q$. This motivates the FBC condition (d). Suppose $\{q_k\}$ and $\{p_k\}$ are two future-ordering chronological sequences exhausting $\hat{q}$.
    If
    $\lim_{k\rightarrow \infty}\phi(q_k) = \tilde{q}$,
    one might conjecture that
    $I^-(\tilde{q}) = \cup_{k} I^-(\phi(q_k)) = \cup_{k} \phi(I^-(q_k)) = \phi(I^-(\hat{q}))$.
    Similarly that if $\lim_{k\rightarrow \infty} \phi(p_k) = \tilde{p}$, then $I^-(\tilde{p}) = \phi(I^-(\hat{q}))$, which implies that
    $\tilde{q} = \tilde{p}$.
    Therefore, the FBC condition (c) appears to be a reasonable assumption to avoid pathological causality conditions. However, this argument based on the past-distinguishing condition remains conjectural, though it can be verified rigorously in the 2-dimensional case. Nevertheless, we can show that for strongly-causal extensions, the FBC condition (d) is automatically satisfied (see Proposition \ref{prop 3.10}).

    Since we are concerned with local extensions in Definition \ref{def 3.6}, removing global hyperbolicity of $(\tilde{M},\tilde{g})$ yields no additional possibilities, as one can always shrink to a small neighbourhood of $\phi(\hat{q})$ to ensure global hyperbolicity. See Lemma \ref{lem 4.11}.

    Furthermore, as noted in \cite{Sb22}, assuming that $\tilde{M}$ and $\phi$ have lower regularity ($\mathrm{C}^k$ for $k\geq 1$) also yields no additional extension possibilities. Therefore, it suffices to assume smoothness in Definitions \ref{def 3.6} and \ref{def 3.7}.
\end{remark}

%%%%%%%%%%%%%%%%%%%%%
%%%%%%%%%%%%%%%%%%%%%
\subsection{Relation between FBC (in)extendibilities}

Intuitively, if $(M,g)$ is locally $\mathrm{C}^{k,\alpha}$ FBC extendible at a future boundary point $\hat{q}$, then $I^-(\hat{q})$ should also be locally $\mathrm{C}^{k,\alpha}$ FBC extendible at $\hat{q}$. The following proposition confirms this expectation.

\begin{proposition}\label{prop 3.9}
    Let $(M,g)$ be a globally hyperbolic spacetime with a future boundary point $\hat{q}$. If $(M,g)$ is locally $\mathrm{C}^{k,\alpha}$ FBC extendible at $\hat{q}$, then $I^-(\hat{q})$ is also locally $\mathrm{C}^{k,\alpha}$ FBC extendible at $q$.
\end{proposition}

\begin{proof}
    Let $(\tilde{M}, \tilde{g}, \phi)$ be a local $\mathrm{C}^{k,\alpha}$ FBC extension of $(M,g)$. Define $\tilde{q} = \phi (\hat{q})$.
    Choose a sufficiently small globally hyperbolic neighbourhood $\tilde{V}$ of $\tilde{q}$. Let
    \begin{align*}
        U = \phi^{-1} (\tilde{V}) \cap I^-(\hat{q}),
        \quad
        \phi_r = \phi|_{U}: U \rightarrow \tilde{V}.
    \end{align*}
    We claim that the triple $(\tilde{V}, \tilde{g}, \phi_r)$ is a local $\mathrm{C}^{k,\alpha}$ FBC extension of $I^-(\hat{q})$ at $\hat{q}$. It suffices to verify condition (b) of Definition \ref{def 3.6}.

    Suppose, for contradiction, that $U$ is not a neighbourhood of $\hat{q}$ in $I^-(\hat{q})$. Let $\{q_k\}_{k\in \mathbb{N}}$ be a future-directed chronological sequence exhausting $\hat{q}$. %in the chronological past.
    Then
    $I(q_k,\hat{q}) \setminus U \neq \emptyset$.
    Thus, we may choose a sequence $\{p_k\}$ such that
    $p_k \in I(q_k, \hat{q}) \setminus U$.
    For each $k$, there exists a smallest index $l_k$ such that
    $p_k \in I^-(q_{l_k})$.
    Define $f: \mathbb{N} \rightarrow \mathbb{N}$ by $f(k) = l_k$.
    Then
    \begin{align*}
        p_k \ll q_{f(k)} \ll p_{f(k)},
        \quad
        f(k) \geq k.
    \end{align*}
    Now define
    $x_n = p_{f^n(0)}$.
    Then $\{ x_n \}$ is a future-oriented chronological sequence exhausting $\hat{q}$.
    However, by condition (d) of Definition \ref{def 3.7},
    $\lim_{n\rightarrow +\infty}\phi(x_n) = \hat{q} \in \tilde{V}$.
    This implies that $x_n \in \phi^{-1}(\tilde{V})\cap I^-(\hat{q}) = U$ for all sufficiently large $n$,
    which contradicts the fact that $x_n = p_{f^n(0)} \in I(q_{f^n(0)}, \hat{q}) \setminus U$.
\end{proof}

%%%%%%%%%%%%%%%%%%%%%
%%%%%%%%%%%%%%%%%%%%%
\subsection{FBC of strongly-causal extension}

\begin{proposition}\label{prop 3.10}
    Let $(M,g)$ be a globally hyperbolic spacetime with a future boundary point $\hat{q}$.
    Suppose
    $(\tilde{M}, \tilde{g}, \phi)$ is a local $\mathrm{C}^{k,\alpha}$ strongly-causal extension of $(M,g)$ along a representative $\{q_k\}$ of $\hat{q}$. Then the FBC condition holds for $\phi$ at $\hat{q}$.
\end{proposition}

\begin{proof}
    Since $Acc(\{ \phi(q_k) \}) \neq \emptyset$, choose $\tilde{q} \in Acc(\{ \phi(q_k) \})$. We show that for any representative $\{p_k\}$ of $\hat{q}$,
    $\lim_{k\rightarrow +\infty} \phi(p_k) = \tilde{q}$.

    By the strong causality condition, for every neighbourhood $\tilde{U}$ of $\tilde{q}$, there exists a causally convex neighbourhood $\tilde{V} \subset \tilde{U}$ of $\tilde{q}$. Since $\phi(q_k)$ converges to $\tilde{q}$, there exists $K$ such that
    $\phi(q_k) \in \tilde{V}$
    for all $k \geq K$.
    Hence, for all $l > k \geq K$,
    \begin{align*}
        J(\phi(q_k), \phi(q_l),\tilde{M}) \subset \tilde{V} \subset \tilde{U}.
    \end{align*}
    Now, for any representative $\{ p_k \}$ of $\hat{q}$, there exist two non-decreasing sequences $\{ m_k \}$ and $\{ n_k \}$ diverging to $+\infty$ such that $p_k \in I(q_{m_k}, q_{n_k})$.
    By the isometry of $\phi$,
    \begin{align*}
        \phi(p_k)
        \in
        \phi\big(I(q_{m_k}, q_{n_k})\big)
        \subset
        I\big( \phi(q_{m_k}), \phi(q_{n_k}), \tilde{M} \big)
        \subset
        J \big( \phi(q_{m_k}), \phi(q_{n_k}), \tilde{M} \big).
    \end{align*}
    Therefore, $\phi(p_k) \in \tilde{U}$ for all sufficiently large $k$ such that $m_k \geq K$. Since $\tilde{U}$ was arbitrary, $\phi(p_k)$ converges to $\tilde{q}$.
\end{proof}

%%%%%%%%%%%%%%%%%%%%%
%%%%%%%%%%%%%%%%%%%%%
\subsection{Localised-chronological-diamond-inclusion of strongly-causal extension}

\begin{definition}\label{def 3.11}
    Let $(M,g)$ be a globally hyperbolic spacetime with $\hat{q}$ as a future boundary point. Suppose $(\tilde{M},\tilde{g}, \phi)$ is a local $\mathrm{C}^{k,\alpha}$ extension along a representative $\{q_k\}$ of $\hat{q}$, and $\tilde{q} \in Acc(\{ \phi (q_k) \})$.
    \begin{enumerate}[label=(\alph*)]
        \item $\phi$ is {\bf chronological-diamond-weakly-inclusive} at $\tilde{q}$ if for every $q \in I^-(\hat{q})$,
              $\phi(I(q,\hat{q})) \subset J(\phi(q), \tilde{q}, \tilde{M})$.
              $\phi$ is {\bf chronological-diamond-inclusive} at $\tilde{q}$ if for every $q \in I^-(\hat{q})$,
              $\phi(I(q,\hat{q})) \subset I(\phi(q), \tilde{q}, \tilde{M})$.
        \item $\phi$ is {\bf localised-chronological-diamond-weakly-inclusive} at $\tilde{q}$ if there exist a neighbourhood $V$ of $\hat{q}$ in $I^-(\hat{q})$ and a neighbourhood $\tilde{V}$ of $\tilde{q}$ in $\tilde{M}$, such that for every $q,q' \in V$, $\phi(I(q,q')) \subset I(\phi(q),\phi(q'),\tilde{V})$ and $\phi(I(q,\hat{q})) \subset J(\phi(q), \tilde{q}, \tilde{V})$.
              $\phi$ is {\bf localised-chronological-diamond-inclusive} at $\tilde{q}$ if for every $q \in V$,
              $\phi(I(q,\hat{q})) \subset I(\phi(q), \tilde{q}, \tilde{V})$.
    \end{enumerate}
\end{definition}

\begin{proposition}\label{prop 3.12}
    Let $(M,g)$ be a globally hyperbolic spacetime with $\hat{q}$ as a future boundary point. Let $(\tilde{M}, \tilde{g}, \phi)$ be a local $\mathrm{C}^{k,\alpha}$ strongly-causal extension at $\hat{q}$, and set
    $\tilde{q} = \phi (\hat{q})$.
    For any neighbourhood $\tilde{V}$ of $\tilde{q}$ in $\tilde{M}$, there exists a neighbourhood $V$ of $\hat{q}$ in $I^-(\hat{q})$ with $\phi(V) \subset \tilde{V}$, such that for any $q \in V$, the following hold:
    \begin{enumerate}[label=(\alph*)]
        \item
              For every $q' \in V$,
              $\phi \big( I(q,q') \big)
                  \subset
                  I (\phi(q), \phi(q'), \tilde{V} )$;
        \item
              Moreover,
              $\phi\big( I(q,\hat{q}) \big)
                  \subset
                  J(\phi(q), \tilde{q}, \tilde{V})$.
    \end{enumerate}
    By definition, $\phi$ is localised-chronological-diamond-weakly-inclusive at $\tilde{q}$.
\end{proposition}

\begin{proof}
    Without loss of generality, we may assume that $\tilde{V}$ is causally convex, since $(\tilde{M},\tilde{g})$ is strongly causal. The existence of $V$ follows from FBC of $\phi$.
    \begin{enumerate}[label=(\alph*)]
        \item
              Since $\tilde{V}$ is causally convex,
              $\phi \big( I(q,q') \big)
                  \subset
                  I(\phi(q), \phi(q'), \tilde{V})$.

        \item
              Let $\{q_k\}$ be a representative of $\hat{q}$.
              For any $q \in M$ such that $\phi(q) \in \tilde{V}$, there exists a sufficiently large $K$ such that for all $k \geq K$,
              $\phi(q) \prec \phi(q_k) \in \tilde{V}$.
              By the causal convexity of $\tilde{V}$, for $k \geq K$ we have
              $\phi(q) \prec_{\tilde{V}} \phi(q_k) \prec_{\tilde{V}} \phi(q_{k+1})$,
              and
              $\phi \big( I(q,q_k) \big)
                  \subset
                  I ( \phi(q), \phi(q_k),\tilde{V} )$.
              Observing that
              $I(q,\hat{q}) = \cup_{k} I(q,q_k)$,
              we conclude
              $\phi \big( I(q,\hat{q}) \big)
                  \subset
                  \cup_{k} I ( \phi(q), \phi(q_k),\tilde{V} )
                  \subset
                  J(\phi(q), \tilde{q}, \tilde{V})$.
    \end{enumerate}
\end{proof}

\begin{remark}
    One may naturally ask whether the reverse inclusion
    in Proposition \ref{prop 3.12} (a) holds. The answer is not obvious, particularly for extensions with low regularity—for example, when $(\tilde{M}, \tilde{g}, \phi)$ is a local $\mathrm{C}^0$ strongly-causal extension.
\end{remark}

%%%%%%%%%%%%%%%%%%%%
%%%%%%%%%%%%%%%%%%%%
%%%%%%%%%%%%%%%%%%%%
\section{Local structure of strongly-causal extension at future boundary}

%%%%%%%%%%%%%%%%%%%%%
%%%%%%%%%%%%%%%%%%%%%
\subsection{Local coordinate system}

\begin{definition}\label{def 4.1}
    Let $p$ be a point in a spacetime $(M,g)$. Let $\{x^0, x^1, \cdots, x^n\}$ be a coordinate system of $(M,g)$ with $p$ as the origin $o$.
    \begin{enumerate}[label=\textbullet]
        \item
              $\{x^0, x^1, \cdots, x^n\}$ is {\bf simple} at $p$
              if $g|_o = \eta = - (\ed x^0)^2 + (\ed x^1)^2 + \cdots + (\ed x^n)^2$.

        \item
              $\{x^0, x^1, \cdots, x^n\}$ is called {\bf $\delta$-rectangular}
              if
              $\vert g_{uv} - \eta_{uv} \vert \leq \delta$.

        \item
              $\{x^0, x^1, \cdots, x^n\}$ is called {\bf $\delta$-simple} at $p$
              if $g|_o=\eta$
              and
              $\vert g_{uv} - \eta_{uv} \vert \leq \delta$.
    \end{enumerate}
\end{definition}

\begin{lemma}\label{lem 4.2}
    Let $g$ be a Lorentzian metric on $\mathbb{R}^{n+1}$ and
    $\vert g_{uv} - \eta_{uv} \vert
        \leq
        \delta
        <
        \frac{1}{n+1}$.
    Let $e_0$ be the point $(1,0,\cdots,0)$ in $\mathbb{R}^{n+1}$.
    Define the Euclidean cone of slope $k$ at $x$ by
    $$
        co_{k}^{\pm}(x)
        =
        \{ y: \pm (y-x)^0
        \geq
        k \sqrt{ \textstyle\sum_{i=1}^n (y-x)^i (y-x)^i }
        \}
    $$
    Define the slopes $k_1,k_2$ by
    $k_1
        =
        \sqrt{ \frac{1 - (n+1)\delta}{1 + (n+1)\delta} }
        <1$,
    $k_2
        =
        k_1^{-1}
        =
        \sqrt{ \frac{1 + (n+1)\delta}{1 - (n+1)\delta} }
        >
        1$.
    See Figure \ref{fig IoC}. Let $I_g(o,e_0)$ and $J_g(o,e_0)$ be the chronological and causal diamonds between $o$ and $e_0$ in $(\mathbb{R}^{n+1},g)$, then
    \begin{align*}
        co_{k_2}^+ (o) \cap co_{k_2}^- (e_0)
        \subset
        I_g(o,e_0)
        \subset
        J_g(o,e_0)
        \subset
        co_{k_1}^+ (o) \cap co_{k_1}^- (e_0).
    \end{align*}
    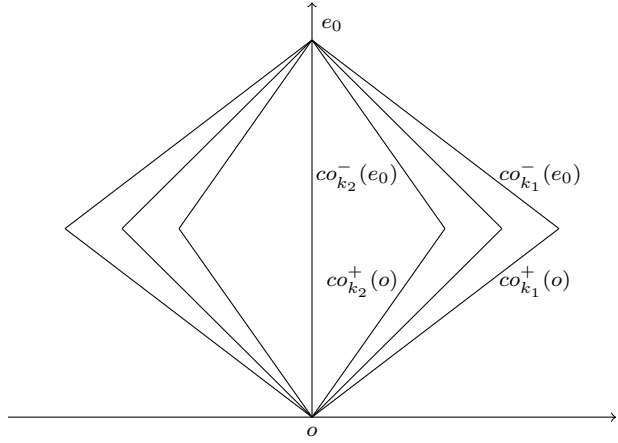
\begin{figure}[h]
        \centering
        \begin{tikzpicture}
            \draw[->] (-4,0) -- (4,0);
            \draw[->] (0,0) -- (0,5.6);
            \node[below] at (0,0) {\footnotesize $o$};
            \node[above right] at (0,5) {\footnotesize $e_0$};
            %%%%%
            % Causal diamond
            %%%%%
            \draw[domain=0:2.5,smooth,variable=\y]
            plot ({\y},{\y});
            \draw[domain=0:2.5,smooth,variable=\y]
            plot ({-\y},{\y});
            \draw[domain=2.5:5,smooth,variable=\y]
            plot ({5-\y},{\y});
            \draw[domain=2.5:5,smooth,variable=\y]
            plot ({-5+\y},{\y});
            %%%%%
            % y= k_1*x, k_1 < 1
            %%%%%
            \draw[domain=0:2.5,smooth,variable=\y]
            plot ({1.3*\y},{\y});
            \draw[domain=0:2.5,smooth,variable=\y]
            plot ({-1.3*\y},{\y});
            \draw[domain=2.5:5,smooth,variable=\y]
            plot ({1.3*(5-\y)},{\y});
            \draw[domain=2.5:5,smooth,variable=\y]
            plot ({1.3*(-5+\y)},{\y});
            \node[right] at ({1.3*1.8},{1.8}) {\footnotesize $co_{k_1}^+(o)$};
            \node[right] at ({1.3*(5-3.2)},{3.2}) {\footnotesize $co_{k_1}^-(e_0)$};
            %%%%%
            % y= k_2*x, k_2 > 1
            %%%%%
            \draw[domain=0:2.5,smooth,variable=\y]
            plot ({0.7*\y},{\y});
            \draw[domain=0:2.5,smooth,variable=\y]
            plot ({-0.7*\y},{\y});
            \draw[domain=2.5:5,smooth,variable=\y]
            plot ({0.7*(5-\y)},{\y});
            \draw[domain=2.5:5,smooth,variable=\y]
            plot ({0.7*(-5+\y)},{\y});
            \node[left] at ({0.7*1.8},{1.8}) {\footnotesize $co_{k_2}^+(o)$};
            \node[left] at ({0.7*(5-3.2)},{3.2}) {\footnotesize $co_{k_2}^-(e_0)$};
        \end{tikzpicture}
        \caption{Inclusions of cones}
        \label{fig IoC}
    \end{figure}
\end{lemma}

\begin{proof}
    By the elementary inequality
    $ ( \sum_{u=0}^n \vert x^u \vert)^2 \leq (n+1) \sum_{u=0}^n (x^u)^2$,
    we obtain
    \begin{align}
        g(x,x)
         &
        \geq
        - \big[ 1 + (n+1)\delta \big]
        \cdot
        \big[ ( x^0 )^2 - k_1^2 \textstyle\sum_{i=1}^n ( x^i )^2 \big],
        \label{eqn 4.1}
        \\
        g(x,x)
         &
        \leq
        - \big[ 1 - (n+1)\delta \big]
        \cdot
        \big[ ( x^0 )^2 - k_2^2 \textstyle\sum_{i=1}^n ( x^i )^2 \big].
        \label{eqn 4.2}
    \end{align}
    The lemma follows from these inequalities.
\end{proof}

%%%%%%%%%%%%%%%%%%%%%
%%%%%%%%%%%%%%%%%%%%%
\subsection{Null cut locus and future horismos}

\begin{definition}\label{def 4.3}
    Let $(M,g)$ be a spacetime. For $p \prec q \in M$, define $\Omega_{p,q}$ as the set of all future-directed causal curves from $p$ to $q$.
    The Lorentz distance $d(p,q)$ is given by
    \begin{align*}
        d(p,q)
        =
        \sup_{\gamma \in \Omega_{p,q}}\{ L[\gamma] \},
    \end{align*}
    where
    $L[\gamma]=\int_I \sqrt{-g(\dot{\gamma}, \dot{\gamma})} \ed t$
    is the length of $\gamma: I \rightarrow M$.
\end{definition}

\begin{definition}[{\bf Future horismos, null cut locus}]\label{def 4.4}
    Let $(M,g)$ be a $\mathrm{C}^2$ globally hyperbolic spacetime and $p \in M$.
    \begin{enumerate}[label=\textbullet]
        \item
              The future horismos $E^+(p)$ is defined by
              $E^+(p) = J^+(p) \setminus I^+(p)$.

        \item
              Let $\gamma_V: [0,a) \rightarrow M$ be an inextendible future-directed null geodesic from $p$ with initial velocity $\dot{\gamma}_V = V$. Define the function $s(V)$ as
              \begin{align}
                  s(V)
                  =
                  \sup \{t\in [0,a): d(p,\gamma_V(t)) = 0 \}.
                  \label{eqn 4.3}
              \end{align}
              We call $\gamma_V(s(V))$ the future null cut point of $p$ along $\gamma_V$.
        \item
              The future null cut locus $C^+_N(p)$ of $p$ consists of all future null cut points of $p$.
        \item
              The open future {\bf null embedding domain} $TE^+_N(p) \subset TC^+(p)$ is
              $TE^+_N(p)
                  =
                  \{ V\in TC^+ (p): s(V) > 1 \}$.
    \end{enumerate}
    The past horismos and past null cut locus can be defined analogously.
\end{definition}

We recall the following property of null cut points (see Theorem 9.15 in \cite{BEE96}).
\begin{proposition}\label{prop 4.5}
    Let $(M,g)$ be a $\mathrm{C}^2$ globally hyperbolic spacetime and $p\in M$. Let $q$ be a future null cut point of $p$ along a future null geodesic $\gamma$. Then at least one of the following holds:
    \begin{enumerate}[label=(\alph*)]
        \item $q$ is the first future conjugate point of $p$ along $\gamma$.
        \item There exist at least two maximal null geodesic segments from $p$ to $q$.
    \end{enumerate}
\end{proposition}

The following proposition characterizes $E^+(p)$ in a globally hyperbolic spacetime.
\begin{proposition}\label{prop 4.6}
    Let $(M,g)$ be a $\mathrm{C}^2$ globally hyperbolic spacetime and $p\in M$. Then the exponential map $\exp$ embeds the future null embedding domain $TE^+_N(p)\setminus\{o\}$ onto its image as a $\mathrm{C}^2$ null hypersurface. Moreover,
    $\overline{\exp(TE^+_N(p))}
        =
        E^+(p)$.
\end{proposition}
\begin{proof}
    The exponential map fails to be an embedding precisely when either (a) or (b) of Proposition \ref{prop 4.5} occurs, establishing the embedding property on $TE^+_N(p)$. We now show that $\overline{\exp(TE^+_N(p))}=E^+(p)$.
    \begin{enumerate}[label=\textbullet]
        \item
              To prove $\exp(TE^+_N(p))\subset E^+(p)$, consider any $V \in \exp(TE^+_N(p))$ and set $q=\exp(V)$. Define the future null geodesic $\gamma_V(t) = \exp(t V)$ and the interval
              $I_V = \gamma^{-1}(I^+(p))$.
              Let
              $t_i = \inf\{ I_V \}$.
              By global hyperbolicity, we have that $t_i >0$ and
              $\gamma_V([0,t_i]) \subset E^+(p)$.
              It suffices to show that $1 \in [0,t_i]$.

              We proceed by contradiction. Suppose $1\notin [0,t_i]$, then $t_i V \in TE^+_N(p)$, so $\exp$ is a local diffeomorphism at $t_i V$. Since $\gamma_V(t_i + \epsilon) \in I^+(p)$, there exists a future timelike vector $V_{\epsilon}$ such that
              $\exp(V_{\epsilon}) = \gamma_V(t_i+\epsilon)$,
              and $\exp(t V_{\epsilon})|_{[0,1]}$ is a maximal timelike geodesic. There exists a sequence $\{\epsilon_k\}$ decreasing to $0$ such that $\{V_{\epsilon_k}\}$ converges to a vector $V'$.
              Then $V'$ is null since $\exp(V') = \gamma_V(1) \notin I^+(p)$. Moreover, by the local diffeomorphism of $\exp$ at $V$, we have
              $V\neq V'$.
              Thus $\gamma_V$ and $\gamma_{V'}$ are two distinct null geodesics from $p$ to $q$, contradicting $V \in TE^+_N(p)$.
        \item
              To prove $E^+(p) \subset \overline{\exp(TE^+_N(p))}$, suppose $q\in E^+(p)$. Then there exists a null vector $V$ such that $\gamma_V(t) = \exp(tV)$ is a maximal null geodesic from $p$ to $q$. Then
              $s(V) \geq 1$,
              where $s(V)$ is defined by Equation \eqref{eqn 4.3}. Hence
              $V \in \overline{TE^+_N(p)}$,
              so $q \in \overline{\exp(TE^+_N(p))}$.
    \end{enumerate}
\end{proof}

%%%%%%%%%%%%%%%%%%%%
%%%%%%%%%%%%%%%%%%%%%
\subsection{Causal/Chronological convexity of domain of dependence}

\begin{lemma}\label{lem 4.7'}
    Let $(M,g)$ be a $\mathrm{C}^0$ spacetime and $S\subset M$ be a closed acausal topological hypersurface. Assume that $q \in \mathcal{D}^{\pm}(S)$ and $\gamma: I= [0,1] \rightarrow M$ is a past/future-directed causal curve emanating from $q$. Then $\gamma^{-1} (\mathcal{D}^{\pm}(S))$ is a connected closed subinterval of $I$ containing $0$.
\end{lemma}
\begin{proof}
    Let $\gamma$ be future-directed and $[0,t_0)$ be the maximal half-open interval such that $\gamma([0,t_0)) \subset \mathcal{D}^+(S)$. Then $\gamma(t_0) \in \mathcal{D}^+(S)$ follows from two observations:
    \begin{enumerate}[label=\textbullet]
        \item
              Every past-directed inextendible causal curve $\tilde{\gamma}: [0,+\infty) \rightarrow M$, which extends $\gamma|_{[0,t_0]}$, intersects $S$ exactly once.
        \item
              $\gamma([0,t_0)) \cap S = \emptyset$: otherwise there exists $t' \in [0,t_0)$ with $\gamma(t') \in S$, then $\tilde{\gamma}|_{(t',+\infty)} \cap S = \emptyset$ by the acausality of $S$, contradicting $\gamma((t',t_0)) \subset \mathcal{D}^+(S)$.
    \end{enumerate}
    The other case follows similarly.
\end{proof}

\begin{proposition}[Causal/Chronological convexity of domain of dependence]\label{prop 4.8}
    Let $(M,g)$ be a $\mathrm{C}^0$ spacetime and $S\subset M$ be a closed acausal topological hypersurface (which is automatically Lipschitz).
    \begin{enumerate}[label=(\alph*)]
        \item
              $\mathcal{D}(S)$, the domain of dependence of $S$, is causally convex.
        \item
              $Int(\mathcal{D}(S))$, the interior of $\mathcal{D}(S)$
              is chronologically convex.
    \end{enumerate}
\end{proposition}
\begin{proof}
    \begin{enumerate}[label=(\alph*)]
        \item
              There are three cases to consider:
              ({i}) $p\prec q \in \mathcal{D}^+(S)$ or
              %({ii}) 
              $p \prec q \in \mathcal{D}^-(S)$;
              ({ii}) $q\in \mathcal{D}^+(S), p\in \mathcal{D}^-(S)$.
              Let $\gamma:I =[0,1] \rightarrow M$ be a past-directed causal curve from $q$ to $p$.
              \begin{enumerate}[label=(\roman*)]
                  \item[(i)]
                        By Lemma \ref{lem 4.7'},
                        $\gamma^{-1}(\mathcal{D}^+(S))
                            =
                            I$.
                        Thus $\gamma(I) \subset \mathcal{D}^+(S)$.
                  \item[(ii)]
                        There exists $t_0\in [0,1]$ such that
                        $\gamma^{-1}(\mathcal{D}^+(S))
                            =
                            [0,t_0]$,
                        $\gamma^{-1}(\mathcal{D}^-(S))
                            =
                            [t_0,1]$,
                        and $\gamma(t_0)$ is the only intersection of $\gamma$ with $S$. Therefore $\gamma(I) \subset \mathcal{D}(S)$.
              \end{enumerate}
        \item
              Similarly, we consider three cases:
              ({i}$'$)
              $p\ll q \in \mathcal{D}^+(S) \cap Int(\mathcal{D}(S))$ or
              $p\ll q \in \mathcal{D}^-(S) \cap Int(\mathcal{D}(S))$;
              ({ii}$'$)
              $q\in \mathcal{D}^+(S) \cap Int(\mathcal{D}(S)), p\in \mathcal{D}^-(S) \cap Int(\mathcal{D}(S))$. Let $\gamma:I =[0,1] \rightarrow M$ be a past-directed chronological curve from $q$ to $p$.
              \begin{enumerate}[label=(\roman*$'$)]
                  \item[(i$'$)]
                        By case ({i}) and the fact that $S$ is a closed Lipschitz hypersurface, we have $\gamma(t) \subset \mathcal{D}^+(S) \setminus S$ for $t\in (0,1)$. Fix $t \in (0,1)$. Then there exists a neighbourhood $U_t$ of $\gamma(t)$ such that
                        $U_t \subset I^-(q) \setminus S$,
                        and for every $p' \in U_t$, there exists a past-directed chronological curve $\gamma_{q,p'}$ from $q$ to $p'$ with
                        $\gamma_{q,p'} \cap S = \emptyset$.
                        Then by observations above, $p' \in \mathcal{D}^+(S)$, which implies
                        $U_t \subset \mathcal{D}^+(S)$.
                  \item[(ii$'$)]
                        Since the case where $q\in S$ or $p\in S$ is covered by cases ({i}$'$), it suffices to consider
                        $q
                            \in
                            \big(\mathcal{D}^+(S) \setminus S \big) \cap Int(\mathcal{D}(S))$ and
                        $p
                            \in
                            \big(\mathcal{D}^-(S) \setminus S \big) \cap Int(\mathcal{D}(S))$.
                        Then there exist neighbourhoods $U_q \subset \mathcal{D}^+(S)$ and $U_p \subset \mathcal{D}^-(S)$ of $q$ and $p$ respectively. By case ({ii}), $I(p,q) \subset Int(\mathcal{D}(S))$.
              \end{enumerate}
    \end{enumerate}
\end{proof}

\begin{remark}\label{rem 4.9}
    If $(M,g)$ is a $\mathrm{C}^2$ spacetime, then it is straightforward to show that $Int(\mathcal{D}(S))$ is causally convex. However, we do not know whether the causal convexity of $Int(\mathcal{D}(S))$ holds for $\mathrm{C}^0$ spacetimes.
\end{remark}

\begin{lemma}\label{lem 4.10}
    Let $(M,g)$ be a $\mathrm{C}^{k,\alpha}$ strongly causal spacetime. For any $\delta>0$ and any $q\in M$, there exists a globally hyperbolic chronologically convex neighbourhood $V$ of $q$, which is covered by a $\delta$-simple coordinate system.
\end{lemma}
\begin{proof}
    Without loss of generality, assume $\delta < \frac{1}{n+1}$. First, there exists a $\delta$-simple coordinate system on a neighbourhood $U_0$ with $q$ as the origin $o$. By the strong causality of $(M,g)$, there exists a causally convex neighbourhood $U_1 \subset U_0$ of $q$. Then $(U_1, g)$ is a $\mathrm{C}^{k,\alpha}$ strongly causal spacetime covered by a $\delta$-simple coordinate system with $p$ as the origin. In this coordinate system, the coordinate disk $D_{\epsilon}^n$
    \begin{align*}
        D_{\epsilon}^n
        =
        \{ v = (0, x^1, \cdots, x^n): \eta (v,v) < \epsilon \}
    \end{align*}
    is acausal in $(U_1, g)$. Define the neighbourhood $V = Int( \mathcal{D}(D_{\epsilon}^n, U_1) ) \subset U_1$.
    By Corollary 5.8 in \cite{Sa16}, $(V,g)$ is globally hyperbolic. Moreover, by Proposition \ref{prop 4.8},
    $V$ is chronologically convex in $(U_1,g)$, hence also in $M$, as $U_1$ is causally convex in $M$.
\end{proof}

\begin{lemma}\label{lem 4.11}
    Let $(M,g)$ be a $\mathrm{C}^0$ spacetime. For any $\delta>0$ and $q \in M$, there exists a globally hyperbolic neighbourhood $V$ of $q$ covered by a $\delta$-simple coordinate system.
\end{lemma}
\begin{proof}
    There exists a $\delta$-simple coordinate system of a neighbourhood $U$ with $q$ as the origin (assume $\delta < \frac{1}{(n+1)}$). Without loss of generality, the coordinate system contains a cylinder
    $R_{\epsilon} = \{ (t,x): \vert t \vert < \epsilon, \vert x \vert < \epsilon \}$.
    Consider the spacelike disc $D^n_{\epsilon/2}$, which is a closed acausal hypersurface in $(R_{\epsilon}, g)$. Define the neighbourhood $V$ of $q$ as
    $V=Int\bigl(\mathcal{D}(D^n_{\epsilon/2},R_{\epsilon})\bigr)$.
    By Corollary 5.8 in \cite{Sa16}, $V$ is globally hyperbolic.
\end{proof}

%%%%%%%%%%%%%%%%%%%%
%%%%%%%%%%%%%%%%%%%%%
\subsection{Chronological-diamond-preservation by null-non-accumulation}

\begin{definition}\label{def 4.12}
    Let $(M,g)$ be a $\mathrm{C}^2$ globally hyperbolic spacetime with a future boundary point $\hat{q}$.  Assume that $(\tilde{M}, \tilde{g}, \phi)$ is a local $\mathrm{C}^{k,\alpha}$ extension of $(M,g)$ at $\hat{q}$ along a future-ordering chronological sequence $\{q_k\}$ exhausting $\hat{q}$.
    \begin{enumerate}[label=(\alph*)]
        \item
              $\phi$ is {\bf null-accumulating} at $\hat{q}$ from $p\in I^-(\hat{q})$ if
              $\phi(\hat{q}) \in Acc( \phi(E^+(p,M)) )$.
        \item
              $\phi$ is {\bf null-non-accumulating} at $\hat{q}$ from $p \in I^-(\hat{q})$ if $\phi(\hat{q}) \notin Acc( \phi(E^+(p,M)) )$.
        \item
              $\phi$ is {\bf locally null-non-accumulating} at $\hat{q}$ if there exists a future-ordering chronological sequence $\{p_k\}\subset I^-(\hat{q})$ exhausting $\hat{q}$,
              such that $\phi$ is null-non-accumulating at $\hat{q}$ from any point in the sequence $\{p_k\}$.
    \end{enumerate}
\end{definition}

\begin{definition}\label{def 4.13}
    Let $(M,g)$ be a $\mathrm{C}^2$ globally hyperbolic spacetime with a future boundary point $\hat{q}$. Let $(\tilde{M}, \tilde{g}, \phi)$ be a local $\mathrm{C}^{k,\alpha}$ strongly-causal extension of $(M,g)$ at $\hat{q}$.
    $\phi$ is {\bf locally chronological-diamond-preserving} at $\hat{q}$ if there exists a neighbourhood $V$ of $\hat{q}$ in $I^-(\hat{q})$ such that for any $p,q\in V$,
    we have
    \begin{align*}
        I(\phi(p), \phi(q), \tilde{M})
        =
        \phi(I(p,q,V)).
    \end{align*}
\end{definition}

\begin{remark}\label{rem 4.14}
    The null-accumulating strongly-causal extension $(\tilde{M}, \tilde{g},\phi)$ of $(M,g)$ at $\hat{q}$ is rather pathological. Heuristically, for a $\mathrm{C}^2$ globally hyperbolic spacetime $(M,g)$ with a null boundary, there will be a null generator of the null boundary that maps to the same point $\phi(\hat{q})$. This implies that the extension fails to be injective at the null boundary. See Example \ref{ex 4.19}. We show that null-accumulation is inconsistent with chronological-diamond-preservation.
\end{remark}

The following lemma concerning the lift of causal curves via an embedding between spacetimes is useful.
\begin{lemma}\label{lem 4.15R}
    Let $(M,g)$ be a globally hyperbolic spacetime. Let $\phi$ be an isometric embedding from $(M,g)$ to a spacetime $(\tilde{M},\tilde{g})$. Let $\tilde{\kappa}: I = [0,1] \rightarrow \tilde{M}$ be a past-directed causal curve from $\tilde{q}=\phi(q)$ to $\tilde{p}_0=\phi(p_0)$. If $\tilde{\kappa}$ can not be lifted to a curve $\kappa$ from $p_0$ to $q$ in $M$, then for all $p \in I^-(q,M)$, we have that
    $\tilde{\kappa}(I) \cap \phi(E^+(p,M)) \neq \emptyset$.
\end{lemma}
\begin{proof}
    Let $I_{s_0} = (s_0,1]$ be the maximal interval in $I$ where $\tilde{\kappa}$ can be lifted via $\phi$ to $\kappa$ in $M$ on $I_{s_0}$. Then we have that
    $\kappa(I_{s_0}) \cap E^+(p,M) \neq \emptyset$,
    otherwise
    $\kappa(I_{s_0}) \subset J(p,q,M)$,
    which implies that $\lim_{s\rightarrow s_0} \kappa(s)$ exists in $J(p,q,M)$ by the compactness of $J(p,q,M)$. The existence of $\lim_{s\rightarrow s_0} \kappa(s)$ contradicts with the maximality of $I_{s_0}$. Hence $\tilde{\kappa}(I) \cap \phi(E^+(p,M))$ contains the nonempty set $\phi\big(\kappa(I_{s_0}) \cap E^+(p,M) \big)$.
\end{proof}

\begin{proposition}\label{prop 4.16}
    Let $(M,g)$ be a $\mathrm{C}^2$ globally hyperbolic spacetime with a future boundary point $\hat{q}$.
    Assume that $(\tilde{M}, \tilde{g}, \phi)$ is a local $\mathrm{C}^{k,\alpha}$ strongly-causal extension of $(M,g)$ at $\hat{q}$, and denote $\phi(\hat{q})$ by $\tilde{q}$. Then there exists a globally hyperbolic chronologically convex neighbourhood $\tilde{V}$ of $\tilde{q}$ such that the following hold:
    \begin{enumerate}[label=(\alph*)]
        \item For any $\phi(p) \ll \phi(q) \in \tilde{V}$, if there exists $p_0 \in I^-(p)$ such that
              $\phi(E^+(p_0,M)) \cap I(\phi(p),\phi(q),\tilde{V}) = \emptyset$,
              then
              $I(\phi(p),\phi(q),\tilde{V})
                  =
                  \phi(I(p,q,M))$.
        \item Furthermore, for any $p \in \phi^{-1}(\tilde{V})$, if there exists $p_0 \in I^-(p)$ such that
              $\phi(E^+(p_0,M)) \cap I(\phi(p),\tilde{q},\tilde{V}) = \emptyset$,
              then
              $I(\phi(p), \tilde{q}, \tilde{M})
                  \subset
                  \phi(I(p,\hat{q}))$.
    \end{enumerate}
\end{proposition}

\begin{proof}
    By Lemma \ref{lem 4.10}, there exists a globally hyperbolic chronologically convex neighbourhood $\tilde{V}$ of $\tilde{q}$.
    \begin{enumerate}[label=(\alph*)]
        \item
              By the global hyperbolicity and chronological convexity of $\tilde{V}$, we have that
              $I(\phi(p),\phi(q),\tilde{M})
                  =
                  I(\phi(p),\phi(q),\tilde{V})$
              is precompact in $\tilde{V}$.

              We also have
              $\phi(I(p,q,M)) \subset I(\phi(p),\phi(q),\tilde{V})$.
              Let $\tilde{\gamma}: I=[0,1] \rightarrow \tilde{M}$ be a past-directed chronological curve from $\phi(q)$ to $\phi(p)$.  It suffices to prove that $\tilde{\gamma}$ can be lifted entirely to a past-directed chronological curve $\kappa$ from $q$ to $p$ in $M$. It follows from Lemma \ref{lem 4.15R}, and
              \begin{align*}
                  \phi(E^+(p_0,M)) \cap \tilde{\kappa}(I)
                  \subset
                  \phi(E^+(p_0,M)) \cap I(\phi(p),\phi(q),\tilde{V})
                  =
                  \emptyset,
              \end{align*}

        \item The statement is trivial if $I(\phi(p),\tilde{q},\tilde{V}) = \emptyset$, so we assume $\phi(p)\ll \tilde{q}$.

              Let $\{ x^0, \dots, x^n\}$ be a $\delta$-rectangular coordinate system at $\tilde{q}$, and consider the coordinate line $l(s) = (s, 0, \dots, 0)$ for $s \in (-\epsilon, 0)$. Then
              $I (\phi(p),\tilde{q},\tilde{V})
                  =
                  \cup_{s\in(-\epsilon,0)}
                  I(\phi(p), l(s), \tilde{V})$.
              For each $s\in (-\epsilon,0)$ there exists $q_s\in I^-(\hat{q})$ such that $\phi(p)\ll l(s) \ll \phi(q_s)$. Consequently, by ({a}),
              \begin{align*}
                  I(\phi(p), l(s), \tilde{V})
                  \subset
                  I(\phi(p),\phi(q_s),\tilde{V})
                  =
                  \phi(I(p,q_s,M))
                  \subset
                  \phi(I(p,\hat{q})).
              \end{align*}
    \end{enumerate}
\end{proof}

\begin{proposition}\label{prop 4.17}
    Let $(M,g)$ be a $\mathrm{C}^2$ globally hyperbolic spacetime with a future boundary point $\hat{q}$. Assume that $(\tilde{M}, \tilde{g}, \phi)$ is a local $\mathrm{C}^{k,\alpha}$ strongly-causal extension of $(M,g)$ at $\hat{q}$. If there exists $p\in I^-(\hat{q})$ such that $\phi$ is null-non-accumulating at $\hat{q}$ from $p$, then $\phi$ is locally chronological-diamond-preserving at $\hat{q}$.
\end{proposition}
\begin{proof}
    By Definition~\ref{def 4.12} of null-non-accumulation, there exists a globally hyperbolic chronological convex neighbourhood $\tilde{U}$ of $\tilde{p}=\phi(\hat{q})$, covered by a $\delta$-simple coordinate system at $p$ with $\delta<\frac{1}{n+1}$, such that
    $\tilde{U} \cap Acc(\phi(E^+(p,M))) = \emptyset$.
    Without loss of generality, we may assume that $\tilde{U}$ also satisfies Proposition~\ref{prop 4.16} by intersecting it with a suitable neighbourhood that does.

    We proceed by contradiction. Suppose $\phi$ is not locally chronological-diamond-preserving at $\hat{q}$. Then, according to Definition~\ref{def 4.13}, and the localised-chronological-diamond-weakly-inclusion of $\phi$ by Proposition \ref{prop 3.12}, there exist a future-ordering sequence $\{r_k\}$ exhausting $\hat{q}$
    and a pair of sequences $\{q_k \in I(r_k,\hat{p}) \}$ and $\{p_k \in I(r_k,\hat{p}) \}$ such that
    $\phi(I(q_k,p_k,M)) \subsetneqq I(\phi(q_k), \phi(p_k), \tilde{U}) \neq \emptyset$.

    By passing to a subsequence, we may also assume that $\{r_k\}$, $\{p_k\}$ and $\{ q_k \}$ are all future-ordering. Since $\phi$ is FBC, all sequences converge to $\tilde{q}$. Without loss of generality, we may further assume that $\phi(r_k)$,
    $\phi(q_k)$,
    $\phi(p_k)$ all lie in $\tilde{U}$.
    Since $\tilde{U}$ is globally hyperbolic and chronological convex,
    $I(\phi(q_k), \phi(p_k), \tilde{U}) = I(\phi(q_k), \phi(p_k), \tilde{M}) \neq \phi(I(q_k,p_k,M))$.
    Proposition~\ref{prop 4.16} yields
    $\phi(E^+(p_0,M)) \cap I(\phi(q_k), \phi(p_k), \tilde{U}) \neq \emptyset$.
    This contradicts
    $\tilde{U} \cap Acc(\phi(E^+(p,M))) = \emptyset$.
    Thus, by contradiction, the corollary is proved.
\end{proof}

%%%%%%%%%%%%%%%%%%%%
%%%%%%%%%%%%%%%%%%%%%
\subsection{Past-chronological-diamond-surjection by null-non-accumulation}

\begin{definition}\label{def 4.18}
    Let $(M, g)$ be a $\mathrm{C}^2$ globally hyperbolic spacetime, and let $\hat{q}$ be a future boundary point of $(M, g)$. Suppose $(\tilde{M}, \tilde{g}, \phi)$ is a local $\mathrm{C}^{k,\alpha}$ strongly-causal extension of $(M, g)$ at $\hat{q}$, and denote $\phi(\hat{q})$ by $\tilde{q}$.
    \begin{enumerate}[label=(\alph*)]
        \item
              $\phi$ is said to be {\bf locally past-chronological-diamond-surjective} at $\hat{q}$ along a future-ordering chronological sequence $\{q_k\} \subset I^-(\hat{q})$ if there exists a future-ordering chronological sequence $\{\tilde{p}_k\} \subset I^-(\tilde{q})$ converging to $\tilde{q}$ such that
              $I(\tilde{p}_k,\tilde{q},\tilde{M}) \subset \phi(I(q_k,\hat{q}))$.
        \item
              $\phi$ is said to be {\bf locally past-chronological-diamond-surjective} at $\hat{q}$ if there exists a future-ordering chronological sequence $\{q_k\}$ exhausting $\hat{q}$ such that $\phi$ is past-chronological-diamond-surjective at $\hat{q}$ along $\{q_k\}$.
    \end{enumerate}
\end{definition}

The following example demonstrates the crucial role of the null-non-accumulating condition in ensuring that an extension is past-chronological-diamond-surjective.
\begin{example}\label{ex 4.19}
    Consider the metric $g$ defined on $U$ by
    \begin{align*}
         &
        g = [\ed u - f(u,x) \ed x] \ed x,
        \quad
        f(u,x) = \frac{\ln (1-u)}{\ln (1-x)} (1-x)^{\frac{\ln (1-u)}{\ln (1-x)} - 1},
        \\
         &
        U
        = \{(x,u)=(x,1-(1-x)^a): x\in(0,1), a\in (2,3)\}\subset \mathbb{R}^2.
    \end{align*}
    The manifold $(U,g)$ exhibits the following properties:
    \begin{enumerate}[label=(\alph*)]
        \item
              $(U,g)$ is a $2$-dimensional smooth Lorentzian manifold with two families of null geodesics:
              \begin{align*}
                  P_{\lambda} = \{x=\lambda\},
                  \quad
                  Q_{\sigma} = \{ 1-u = (1-x)^{\sigma}\},
                  \quad
                  \lambda \in (0,1),
                  \quad
                  \sigma \in (2,3).
              \end{align*}
        \item
              In the double null coordinate system $(\lambda, \sigma) \in (0,1)\times(2,3)$,
              \begin{align*}
                   &
                  f(u,x)
                  =
                  \sigma (1-\lambda)^{\sigma-1},
                  \\
                   &
                  \ed x = \ed \lambda,
                  \quad
                  \ed u
                  =
                  \sigma (1-\lambda)^{\sigma-1} \ed \lambda
                  -
                  (1-\lambda)^{\sigma}\ln (1-\lambda) \ed \sigma,
                  \\
                   &
                  g = -(1-\lambda)^{\sigma} \ln(1-\lambda) \ed \sigma \ed \lambda.
              \end{align*}
              The spacetime $(U,g)$ is globally hyperbolic.
        \item
              The function $f(u,x)$ extends continuously to $\tilde{U} = \mathbb{R}_{>0} \times \mathbb{R}$:
              \begin{align*}
                  \tilde{f}(u,x)
                  =
                  \left\{
                  \begin{aligned}
                       &
                      f(u,x)
                      \quad
                      (u,x) \in U,
                      \\
                       &
                      2 (1-x),
                      \quad
                      1-u \geq (1-x)^2, x\in (0,1),
                      \\
                       &
                      3 (1-x)^2,
                      \quad
                      1-u \leq (1-x)^3, x\in (0,1),
                      \\
                       &
                      0,
                      \quad
                      x \in [1,+\infty).
                  \end{aligned}
                  \right.
              \end{align*}
        \item
              The metric $g$ extends continuously to a metric $\tilde{g}$ on $\tilde{U}$:
              $\tilde{g} = [\ed u - \tilde{f}(u,x) \ed x] \ed x$.
              A direct verification shows that $(\tilde{U},\tilde{g})$ is strongly causal.
        \item
              Let $\phi$ be the inclusion of $U$ into $\tilde{U}$. Then $(\tilde{U},\tilde{g},\phi)$ is a $\mathrm{C}^0$ strongly-causal extension of $(U,g)$ at $(1,1)$.
        \item
              The entire future null boundary $\{(\lambda, \sigma): \lambda =1, \sigma \in (2,3) \}$ maps to the single point $(1,1)$. Consequently, $\phi$ is null-accumulating, and it is not past-chronological-diamond-surjective at the future boundary point $(\lambda,\sigma) = (1,3)$.
        \item
              $U$ is contained in the bubble region (in the sense of \cite{CG12}) of $(1,1)$ in $(\tilde{U},g)$.
              $U \cap I^-((1,1), \tilde{U}) = \emptyset$.
    \end{enumerate}
\end{example}

\begin{proposition}\label{prop 4.20}
    Let $(M,g)$ be a $\mathrm{C}^2$ globally hyperbolic spacetime with future boundary point $\hat{q}$. Suppose $(\tilde{M}, \tilde{g},\phi)$ is a local $\mathrm{C}^{k,\alpha}$ strongly-causal extension of $(M,g)$ at $\hat{q}$, and denote $\phi(\hat{q})$ by $\tilde{q}$.
    \begin{enumerate}[label=(\alph*)]
        \item
              If $\phi$ is null-non-accumulating at $\hat{q}$ from a point $p\in I^-(\hat{q})$, then there exists a future-ordering chronological sequence $\{q_k\} \subset I^-(\hat{q})$ escaping every compact set, such that $\phi$ is locally past-chronological-diamond-surjective at $\hat{q}$ along $\{q_k\}$.
        \item
              If $\phi$ is locally null-non-accumulating at $\hat{q}$, then $\phi$ is locally past-chronological-diamond-surjective at $\hat{q}$.
    \end{enumerate}
\end{proposition}
\begin{proof}
    \begin{enumerate}[label=(\alph*)]
        \item
              Since $\phi$ is null-non-accumulating at $\hat{q}$ from $p$, Proposition \ref{prop 4.16} ensures the existence of a future-ordered chronological sequence $\{p_k\} \subset I^+(p)$ exhausting $\hat{q}$ such that
              $I(\tilde{p}_k,\tilde{p}_l,\tilde{M})
                  =
                  \phi(I(p_k,p_l,M))$,
              $\tilde{p}_k = \phi(p_k)$.
              Choose a future-ordering chronological sequence $\{\tilde{q}_k\}\subset I^-(\tilde{q})$ converging to $\tilde{q}$. Let $\tilde{\gamma}_k: [0,1] \rightarrow \tilde{M}$ be a past-directed chronological curve from $\tilde{p}_k$ to $\tilde{q}_k$. Since both $\{\tilde{p}_k\}$ and $\{\tilde{q}_k\}$ converge to $\tilde{q}$, $\{\tilde{\gamma}_k\}$ converges to $\tilde{q}$. The null-non-accumulation condition $\tilde{q} \notin \phi(E_+(p,M))$ implies that for sufficiently large $k>K$,
              $\tilde{\gamma}_k([0,1]) \cap \phi(E_+(p,M)) = \emptyset$
              By Lemma \ref{lem 4.15R}, $\tilde{\gamma}_k$ can be lifted to a past-directed chronological curve $\gamma_k$ from $p_k$ to $\gamma_k(1)$ in $I^+(p,p_k,M)$. Define
              $q_k = \gamma_k(1) = \phi^{-1} (\tilde{q}_k) \in I(p,\hat{q})$.
              Thus, $\{\tilde{q}_k\}_{k>K} \subset \phi(I(p,\hat{q}))$.
              By Proposition \ref{prop 4.16}, for sufficiently large $k>K$,
              $I(\tilde{q}_k, \tilde{q}_l,\tilde{M})
                  =
                  \phi(I(q_k, q_l,M))$
              and
              $I(\tilde{q}_k, \tilde{q},\tilde{M})
                  =
                  \cup_{l \geq k} I(\tilde{q}_k, \tilde{q}_l,\tilde{M})
                  \subset
                  \phi(I(q_k,\hat{q}))$.
              Therefore, $\phi$ is locally past-chronological-diamond-surjective at $\hat{q}$ along $\{q_k\}_{k>K}$. Since $\phi(q_k) \rightarrow \tilde{q} \notin \phi(M)$, $\{q_k\}$ is future-ordering, chronological, and escapes every compact set
        \item
              Let $\{r_k\}$ be a future-ordering chronological sequence exhausting $\hat{q}$, with $\phi$ null-non-accumulating at $\hat{q}$ from each $r_k$. By a standard diagonal argument, there exists a future-ordering chronological sequence $\{q_k \in I(r_k,\hat{q})\}$ that exhausts $\hat{q}$, such that $\{\tilde{q}_k = \phi(q_k) \ll \tilde{q}\}$ is future-ordering, chronological, and converges to $\tilde{q}$,
              $I(\tilde{q}_k,\tilde{q}, \tilde{M}) \subset \phi(I(q_k,\hat{q}))$.
              Hence, $\phi$ is locally past-chronological-diamond-surjective at $\hat{q}$.
    \end{enumerate}
\end{proof}

%%%%%%%%%%%%%%%%%%%%
%%%%%%%%%%%%%%%%%%%%%
\subsection{Condition for null-non-accumulation: Compact future horismos}

\begin{proposition}\label{prop 4.21}
    Let $(M,g)$ be a $\mathrm{C}^2$ globally hyperbolic spacetime with $\hat{q}$ a future boundary point of $(M,g)$. If $E^+(p,M)$ is compact for $p\in I^-(\hat{q})$, then for every local extension $(\tilde{M},\tilde{g},\phi)$ of $(M,g)$ at $\hat{q}$, $\phi$ is null-non-accumulating at $\hat{q}$ from $p$.
\end{proposition}
\begin{proof}
    Since $\phi(E^+(p,M))$ is compact and $\phi(\hat{q}) \notin \phi(E^+(p,M))$, it follows that
    $\phi(\hat{q}) \notin Acc(\phi(E^+(p,M)))$.
\end{proof}

\begin{example}[Misner spacetime]\label{ex 4.22}
    Let $L_{\phi}$ be the Lorentz boost of hyperbolic angle $\phi$ on the two-dimensional Minkowski spacetime $(\mathbb{M}^{2},\eta)$, defined by
    \begin{align*}
        L_{\phi}: (l \cosh(\theta), l \sinh(\theta)) \mapsto (l\cosh(\theta+\phi), l \sinh(\theta+\phi)).
    \end{align*}
    The Misner spacetime is the quotient spacetime
    $(M,\eta) = (I^-(o)/\mathbb{Z} L_{\phi}, \eta)$.
    One can directly verify that for any $p \in M$, the set $E_+(p,M)$ is compact.
\end{example}

\begin{example}\label{ex 4.23}
    Consider the spatially spherical FLRW spacetime $(M=\mathbb{R}_{-}\times \mathbb{S}^n, g)$, where
    $g= -\ed t^2 + t^2 \ringg$
    and $\ringg$ denotes the standard round metric on the sphere. A direct computation confirms that for every $p\in M$, the set $E^+(p,M)$ is compact.
\end{example}

%%%%%%%%%%%%%%%%%%%%
%%%%%%%%%%%%%%%%%%%%%
\subsection{Condition for null-non-accumulation: $\mathrm{C}^{0,1}$ strongly-causal extension}

\begin{lemma}\label{lem 4.24}
    Let $g$ be a $\mathrm{C}^{0,1}$ Lorentzian metric on $U \subset \mathbb{R}^{n+1}$. Suppose that
    \begin{align*}
        \lim_{y\rightarrow x} \frac{\vert g_{\mu\nu}(x) - g_{\mu\nu}(y) \vert}{\vert x - y \vert}
        \leq
        \Delta,
        \quad
        \vert g_{\mu\nu}(x) - \eta_{\mu\nu} \vert
        \leq
        \delta < \frac{1}{n+1}.
    \end{align*}
    Let $\gamma$ be a future-directed null geodesic satisfying the geodesic equation
    $\ddot{\gamma}^{\kappa} + \Gamma_{\mu\nu}^{\kappa} \dot{\gamma}^{\mu} \dot{\gamma}^{\nu} =0$,
    where $\Gamma_{\mu\nu}^{\kappa} = \frac{1}{2}(g^{-1})^{\kappa\lambda}(\partial_{\mu} g_{\nu \lambda} + \partial_{\nu} g_{\mu\lambda} - \partial_{\lambda} g_{\mu\nu})$ exists along $\gamma$. Then there exists a constant $C$ depending on $n$ such that for any
    $\vert s \vert
        \leq
        \frac{1}{C\Delta \cdot \vert \dot{\gamma}^0(0) \vert}$,
    we have that
    \begin{align*}
        \vert \dot{\gamma}^{\kappa}(s) - \dot{\gamma}^{\kappa}(0) \vert
        \leq
        C \Delta s \vert \dot{\gamma}^0(0) \vert^2,
        \quad
        \vert \gamma^{\kappa}(s) - \sigma^{\kappa}(s) \vert
        \leq
        C \Delta s^2 \vert \dot{\gamma}^0(0) \vert^2,
    \end{align*}
    where
    $\sigma^{\kappa}(s) = \gamma^{\kappa}(0) + \dot{\gamma}^{\kappa}(0) s$.
\end{lemma}

\begin{proof}
    Throughout this proof, we denote constants depending on $n$ by $c(n)$. There exists a constant $c(n)$ such that
    \begin{align*}
        \vert \Gamma_{\mu\nu}^{\kappa} \vert \leq c(n) \Delta,
        \quad
        \vert \eta_{\mu\nu} \dot{\gamma}^{\mu} \dot{\gamma}^{\nu} \vert
        <
        c(n)\delta (\dot{\gamma}^0)^2,
        \quad
        \max_{\kappa} \{ \vert \dot{\gamma}^{\kappa} \vert \}
        \leq
        c(n) \vert \dot{\gamma}^0 \vert.
    \end{align*}
    Then
    \begin{align*}
        \vert \ddot{\gamma}^k \vert
        \leq
        c(n) \Delta \vert \dot{\gamma}^0 \vert^2
        \quad
        \Rightarrow
        \quad
        \vert \frac{\ed}{\ed s} \frac{1}{\dot{\gamma}^0} \vert
        \leq
        c(n) \Delta
        \quad
        \Rightarrow
        \quad
        \vert \frac{1}{\dot{\gamma}^0(s)} - \frac{1}{\dot{\gamma}^0(0)} \vert
        \leq
        c(n)\Delta s.
    \end{align*}
    There exists a constant $c_0$ depending on $n$ such that for $\vert s \vert \leq \frac{1}{c_0 \Delta \cdot \vert \dot{\gamma}^0(0) \vert}$,
    \begin{align*}
        \vert \dot{\gamma}^0(s) - \dot{\gamma}^0(0) \vert
        \leq
        c_0 \Delta s \vert \dot{\gamma}^0(0) \vert^2,
        \quad
        \vert \dot{\gamma}^0(s) \vert \leq 2 \vert \dot{\gamma}^0(0) \vert.
    \end{align*}
    Then for $\vert s \vert \leq \frac{1}{c_0 \Delta \cdot \vert \dot{\gamma}^0(0) \vert}$,
    $\vert \ddot{\gamma}^{\kappa} \vert \leq c(n)\Delta \vert \dot{\gamma}^0(0) \vert^2$.
    Integrating this inequality completes the proof.
\end{proof}

\begin{lemma}\label{lem 4.25}
    Let $g$ be a $\mathrm{C}^{0,1}$ Lorentzian metric on $U$. Suppose that $\gamma: [t_0,t_1) \rightarrow U$ is a future-directed non-spacelike geodesic satisfying the geodesic equation
    $\ddot{\gamma}^{\kappa} + \Gamma_{\mu\nu}^{\kappa} \dot{\gamma}^{\mu} \dot{\gamma}^{\nu} =0$,
    where $\Gamma_{\mu\nu}^{\kappa}$ exists along $\gamma$.
    Moreover, assume that the differential $\ed g_{\mu\nu}$ exists along $\gamma([t_0,t_1))$.
    Suppose that
    $\lim_{t\rightarrow t_1} \gamma(t) = q \in U$.
    Then $\gamma$ is Lipschitz-continuously differentiable at $t_1$ with future timelike tangent vector $\dot{\gamma}(t_1)$.
\end{lemma}

\begin{proof}
    Let $\{x^0,x^1,\cdots,x^n\}$ be a $\delta$-simple
    coordinate system at $q$ with $\delta <\frac{1}{100n^2}$. Define the small rectangle
    $R_{l}= \{x: \vert x^i \vert < l \}$.
    Assume that $\tilde{\gamma}([t_1 - \epsilon, t_1]) \subset R_l$.
    For $t\in [t_1-\epsilon, t_1)$, since $\dot{\gamma}$ is future timelike, we have
    \begin{align*}
        \dot{\gamma}^0(t) > 0,
        \quad
        \gamma^0(t_1-\epsilon) > -l,
        \quad
        \gamma^0 (t) \in [\gamma^0 (t_1-\epsilon), 0],
        \quad
        \vert \dot{\gamma}^i (t) \vert    <    2 \dot{\gamma}^0 (t).
    \end{align*}
    Therefore, by the geodesic equation, there exists a constant $C$ depending on the Lipschitz constant of $g_{\mu\nu}$ and $n$ such that
    \begin{align*}
        \vert \frac{\ed}{\ed t} \dot{\gamma}^0 \vert
        <
        C \vert \dot{\gamma}^0 \vert^2
        \quad
        \Rightarrow
        \quad
        \vert \frac{\ed}{\ed t} \frac{1}{\dot{\gamma}^0} \vert
        <
        C.
    \end{align*}
    For any $t_1 - \epsilon < t' < t < t_1$,
    \begin{align*}
        \frac{1}{\dot{\gamma}^0(t')}
        <
        \frac{1}{\dot{\gamma}^0(t)} + C(t - t')
        \quad
        \Rightarrow
        \quad
        \dot{\gamma}^0(t')
        >
        \frac{\dot{\gamma}^0(t)}{1 + C \dot{\gamma}^0(t) (t-t') }.
    \end{align*}
    Integrating $\dot{\tilde{\gamma}}^0$ yields
    \begin{align*}
        \gamma^0(t) - \gamma^0(t_1 - \epsilon)
         & >
        \int_{t_1 - \epsilon}^{t} \frac{\dot{\gamma}^0(t)}{1 + C \dot{\gamma}^0(t) (t-t') } \ed t'
        =
        - \frac{1}{C} \log [ 1 + C \dot{\gamma}^0(t) (t -t')] \big|_{t_1 - \epsilon}^{t}
        \\
         & =
        \frac{1}{C} \log [ 1 + C \dot{\gamma}^0(t) (t -t_1 + \epsilon)].
    \end{align*}
    There exists $N > 0$ such that
    $\limsup_{t \rightarrow t_1^-} \dot{\gamma}^0(t)
        <
        N$,
    because
    \begin{align*}
         &
        1 + C \dot{\gamma}^0(t) (t -t_1 + \epsilon)
        <
        \exp [ C (\gamma^0(t) - \gamma^0(t_1 - \epsilon) ) ]
        \\
        \Rightarrow
        \quad
         &
        \dot{\gamma}^0(t)
        \leq
        \frac{
            \exp [ C (\gamma^0(t) - \gamma^0(t_1 - \epsilon) ) ] - 1
        }{
            C ( t - t_1 + \epsilon )
        }
        \leq
        \frac{
            \exp(C\cdot l) - 1
        }{
            C ( t - t_1 + \epsilon )
        }.
    \end{align*}
    For sufficiently small $\vert t - t_1 \vert$,
    $\vert \frac{\ed}{\ed t} \dot{\gamma}^0 (t) \vert
        <
        C N^2$,
    and similarly for the other components.
    Therefore $\dot{\gamma}(t)$ is uniformly bounded on $[t_1 - \epsilon, t_1)$ and extends Lipschitz-continuously to $t_1$. %The vector 
    $\dot{\gamma}(t_1)$ is timelike by the conservation of $\tilde{g} (\dot{\gamma}, \dot{\gamma})$ from the geodesic equation.
\end{proof}

\begin{proposition}\label{prop 4.26}
    Let $(M,g)$ be a $\mathrm{C}^2$ globally hyperbolic spacetime with a future boundary point $\hat{q}$. Suppose there exists a future-directed chronological geodesic $\gamma:[t_0,0) \rightarrow M$ parametrized by arc length that converges to $\hat{q}$. Then any $\mathrm{C}^{0,1}$ strongly-causal extension $(\tilde{M},\tilde{g},\phi)$ of $(M,g)$ at $\hat{q}$ is null-non-accumulating at $\hat{q}$ from a sufficiently short segment $\gamma|_{(-\epsilon,0)}$.
\end{proposition}
\begin{proof}
    Define $\tilde{\gamma} = \phi \circ \gamma: [t_0, 0) \rightarrow \tilde{M}$.
    Let $\tilde{q}$ denote $\phi(\hat{q})$. By Proposition \ref{prop 3.10}, $\tilde{\gamma}$ converges to $\tilde{q}$. According to Lemma \ref{lem 4.25}, $\tilde{\gamma}$ is Lipschitz-continuously differentiable at $t=0$ with the future timelike tangent vector $\dot{\gamma}(0)$. Choose a $\delta$-simple coordinate system $\{x^0, \cdots, x^n\}$ at $\tilde{q}$ of $(\tilde{M},\tilde{g})$, where $\delta<\frac{1}{n+1}$, such that
    $\dot{\tilde{\gamma}}(0) = e_0 = (1,0,\cdots,0)$.
    Then for sufficiently small $\epsilon$ and $- \epsilon < t < 0$,
    \begin{align*}
        \vert \tilde{\gamma}^{\kappa}(t) - \delta_0^{\kappa} t \vert
        =
        O(t^2).
    \end{align*}
    Let $\tau_t$ be a future-directed null geodesic in $M$ emanating from $\tau_t(0) = \gamma(t)$ and define
    $\tilde{\tau}_t = \phi \circ \tau_t$.
    Without loss of generality, assume that
    \begin{align*}
        \dot{\tilde{\tau}}_t^0(0) = 1,
        \quad
        \dot{\tilde{\tau}}_t^i(0) = v^i,
        \quad
        \vert v \vert = 1 + O(t).
    \end{align*}
    Define the line $\sigma_{\tau_t}$ by $\sigma_{\tau_t}(s) = \tau_t(0) + \dot{\tilde{\tau}}_t(0) \cdot s$. Then
    \begin{align*}
         &
        \sigma_{\tau_t}^0(s)
        =
        \tilde{\tau}_t^0(0) + \dot{\tilde{\tau}}_t^0(0) \cdot s
        =
        t + s + O(t^2),
        \\
         &
        \sigma_{\tau_t}^i(s)
        =
        \tilde{\tau}_t^i(0) + \dot{\tilde{\tau}}_t^i(0) \cdot s
        =
        v^i \cdot s + O(t^2).
    \end{align*}
    Let $\Delta$ be the Lipschitz constant of the metric components $g_{\mu\nu}$. Then by Lemma \ref{lem 4.24}, there exists a constant $C$ depending on $n$ such that for any
    $\vert s \vert \leq \frac{1}{C\Delta}$,
    we have
    \begin{align*}
        \vert \tilde{\tau}_t^{\kappa}(s) - \sigma_{\tau_t}^{\kappa}(s) \vert
        \leq
        C \Delta s^2.
    \end{align*}
    Therefore for $\vert s \vert \leq \frac{1}{C\Delta}$,
    \begin{align*}
        \tilde{\tau}_t^0(s)
        =
        t + s + O(t^2 + s^2),
        \quad
        \tilde{\tau}_t^i(s)
        =
        v^i \cdot s + O(t^2+s^2).
    \end{align*}
    Solving the equation $\tilde{\tau}_t^0(s) =0$ and noting that $\tilde{\tau}_t^0$ is strictly increasing, we obtain
    \begin{align*}
        s= t + O(t^2),
        \quad
        \tilde{\tau}_t^i(s) = v^i \cdot t + O(t^2).
    \end{align*}
    For sufficiently small $t$, $\tilde{\tau}_t$ does not pass through $\tilde{q}$. Therefore by Proposition \ref{prop 4.6}, for sufficiently small $t$,
    $\tilde{q} \notin Acc(E^+(\gamma(t), M))$.
    Hence, $\phi$ is null-non-accumulating at $\hat{q}$ from $\gamma|_{(-\epsilon,0)}$ for sufficiently small $\epsilon$.
\end{proof}

%%%%%%%%%%%%%%%%%%%%%
%%%%%%%%%%%%%%%%%%%%%
%%%%%%%%%%%%%%%%%%%%%
\section{VDR asymptote}

This section addresses the asymptotic behavior of volumes of chronological and causal diamonds.
\begin{problem}\label{prob 5.1}
Let $p$ be a point in a $\mathrm{C}^{k,\alpha}$ globally hyperbolic spacetime $(M,g)$. Study the asymptotic behavior of the volume of the chronological or causal diamond $I(p,q)$ or $J(p,q)$ with respect to the Lorentz distance $d(p,q)$ as $q$ approaches $p$ within a compact range of timelike directions.
\end{problem}

The above problem formulation aims to motivate the subsequent definitions and results, rather than provide a rigorous statement.

%%%%%%%%%%%%%%%%%%%%%
%%%%%%%%%%%%%%%%%%%%%
\subsection{Distance and volume estimates in simple coordinate system}

\begin{definition}\label{def 5.2}
    Let $\{x^0, x^1, \cdots, x^n\}$ be a simple coordinate system at $p$. The coordinate future(past) timelike cone of hyperbolic angle $\varphi$ in $\mathbb{R}^{n+1}$, denoted by $O_{\varphi}^{+(-)}$, is defined by
    \begin{align*}
        O_{\varphi}^{+(-)}
        =
        \{\  v:
        \text{ $v$ is a future(past) timelike vector in $(\mathbb{R}^{n+1}, \eta)$}, &
        \\
        \text{$\vert \eta ( v, \partial_0 ) \vert
                \leq
                \cosh \varphi \cdot \sqrt{\vert \eta ( v, v ) \vert }
        $}                                                                           &
        \}.
    \end{align*}
\end{definition}

\begin{lemma}\label{lem 5.3}
    Let $g$ be a Lorentzian metric in $\mathbb{R}^{n+1}$ and $\vert g_{uv} - \eta_{uv} \vert \leq \delta < \frac{1}{n+1}$. Let $e_0$ be the point $(1,0,\cdots,0)$ in $\mathbb{R}^{n+1}$. There exists a constant $C$ such that:
    Let $d_g(o,e_0)$ be the distance between $o$ and $e_0$ in $(\mathbb{R}^{n+1},g)$, then
    $\vert d_g(o,e_0) - 1 \vert \leq C \delta$.
\end{lemma}

\begin{proof}
    \begin{enumerate}[label=(\roman*)]
        \item
              $d_g(o,e_0) \geq 1 - C \delta$: Let $\gamma(t) = t e_0$ be the timelike curve. Then
              \begin{align*}
                  d_g(o,e_0)
                  \geq
                  L_g [ \gamma|_{[0,1]} ]
                  =
                  \int_0^1 \sqrt{ g(e_0, e_0) } \ed t
                  \geq
                  \int_0^1 \sqrt{ 1 - \delta } \ed t
                  \geq
                  1 - \frac{\delta}{2}.
              \end{align*}

        \item
              $d_g(o,e_0) \leq 1 + C \delta$: For any timelike curve $\gamma$ from $o$ to $e_0$. By \eqref{eqn 4.1}, we have
              \begin{align*}
                  -g(\dot{\gamma}, \dot{\gamma})
                  \leq
                  [1 + (n+1)\delta] \dot{\gamma}^0 \dot{\gamma}^0
                  - [1 - (n+1)\delta] \sum_{i=1}^n \dot{\gamma}^i \dot{\gamma}^i.
              \end{align*}
              Then
              $L_g(\gamma)
                  =
                  \int_{\gamma} \sqrt{-g(\dot{\gamma}, \dot{\gamma})} \ed t
                  \leq
                  \sqrt{1+(n+1)\delta} \int_{\gamma} \dot{\gamma}^0 \ed t
                  \leq
                  1+\frac{(n+1)}{2} \delta$.
    \end{enumerate}
\end{proof}

\begin{lemma}\label{lem 5.4}
    Let $g$ be a Lorentzian metric in $\mathbb{R}^{n+1}$ and $\vert g_{uv} - \eta_{uv} \vert \leq \delta < \frac{1}{100n^2}$. Let $e_0$ be the point $(1,0,\cdots,0)$ in $\mathbb{R}^{n+1}$. There exists a constant $C$ such that:
    Let $I_g(o,e_0)$ and $J_g(o,e_0)$ be the chronological and causal diamonds between $o$ and $e_0$. Let $\vert I_g(o,e_0) \vert_g$ and $\vert J_g(o,e_0) \vert_g$ be their volumes in $(\mathbb{R}^{n+1},g)$. Similarly for $\vert I_{\eta}(o,e_0) \vert_{\eta}$ and $\vert J_{\eta}(o,e_0) \vert_{\eta}$. Then
    \begin{align*}
         &
        \big\vert \frac{\vert I_g(o,e_0) \vert_g}{\vert I_{\eta}(o,e_0)\vert_{\eta}} - 1   \big\vert \leq C \delta,
        \quad
        \big\vert \frac{\vert J_g(o,e_0) \vert_g}{\vert J_{\eta}(o,e_0) \vert_{\eta}} -1  \big\vert \leq C \delta.
    \end{align*}
\end{lemma}

\begin{proof}
    We compare the volume forms $\dvol_g$ and $\dvol_{\eta}$: for $\delta < \frac{1}{100n^2}$
    \begin{align*}
        (1 - 2\delta)^{n+1} \dvol_{\eta}
        \leq
        \dvol_g
        \leq
        (1+2\delta)^{n+1} \dvol_{\eta}.
    \end{align*}
    By Lemma \ref{lem 4.2},
    $co_{k_2}^+ (o) \cap co_{k_2}^- (e_0)
        \subset
        I_g(o,e_0)
        \subset
        J_g(o,e_0)
        \subset
        co_{k_1}^+ (o) \cap co_{k_1}^- (e_0)$.
    Therefore,
    \begin{align*}
        (1-2\delta)^{n+1} \vert co_{k_2}^+ (o) \cap co_{k_2}^- (e_0) \vert_{\eta}
         &
        \leq
        \vert I_g(o,e_0) \vert_{g}
        \\
         &
        \leq
        \vert J_g(o,e_0) \vert_{g}
        \leq
        (1+2\delta)^{n+1} \vert co_{k_1}^+ (o) \cap co_{k_1}^- (e_0) \vert_{\eta}.
    \end{align*}
    We now estimate the volumes of $\vert I_g(o,e_0) \vert_g$ and $\vert J_g(o,e_0) \vert_g$.
    \begin{enumerate}[label=(\roman*)]
        \item
              $\vert I_g(o,e_0) \vert_g \geq (1-C\delta)\vert I_\eta(o,e_0) \vert_{\eta}$:
              Observe that
              \begin{align*}
                   &
                  \vert I_g(o,e_0) \vert_g
                  \geq
                  (1 - 2\delta)^{n+1} (k_2^{-1})^n \vert co_{1}^+(o) \cap co_{1}^-(e_0) \vert_{\eta}.
                  \\
                   &
                  k_2^{-1}
                  =
                  k_1
                  >
                  1 - (n+1) \delta,
                  \quad
                  co_{1}^+(o) \cap co_{1}^-(e_0)
                  =
                  I_{\eta}(o,e_0).
              \end{align*}
              Then
              \begin{align*}
                  \vert I_g(o,e_0) \vert_g
                   &
                  \geq
                  (1 - 2\delta)^{n+1} \big( 1 - (n+1) \delta \big)^n
                  \cdot
                  \vert I_{\eta}(o,e_0) \vert_{\eta}
                  \\
                   &
                  \geq
                  \big( 1 -  (n+2)(n+1) \delta \big)
                  \cdot
                  \vert I_{\eta}(o,e_0) \vert_{\eta}.
              \end{align*}

        \item
              $\vert I_g(o,e_0) \vert_g \leq (1 + C\delta) \vert I_\eta(o,e_0) \vert_{\eta} $:
              Observe that
              \begin{align*}
                  \vert I_g(o,e_0) \vert_g
                  \leq
                  (1 + 2\delta)^{n+1} (k_1^{-1})^n \vert I_{\eta}(o,e_0) \vert_{\eta}
              \end{align*}
              Note that for $\delta < \frac{1}{100n^2}$,
              \begin{align*}
                   &
                  k_1^{-1}
                  =
                  k_2
                  <
                  1 + 2(n+1) \delta,
                  \\
                   &
                  \big( 1 + 2(n+1) \delta \big)^{2n+1}
                  \left\{
                  \begin{aligned}
                       &
                      > (1 + 2\delta)^{n+1} \cdot (k_1^{-1})^n,
                      \\
                       &
                      \leq 1 + (2n+1) \big( 1+ \frac{n+1}{50n^2} \big)^{2n} (2n+2)\delta.
                  \end{aligned}
                  \right.
              \end{align*}
              Then
              \begin{align*}
                  \vert I_g(o,e_0) \vert_g
                   &
                  \leq
                  \Big( 1 + (2n+1) \big( 1+ \frac{n+1}{50n^2} \big)^{2n} \cdot (2n+2)\delta \Big)
                  \vert I_{\eta}(o,e_0) \vert_{\eta}
              \end{align*}
    \end{enumerate}
\end{proof}

%%%%%%%%%%%%%%%%%%%%%
%%%%%%%%%%%%%%%%%%%%%
\subsection{VDR asymptote in $\mathrm{C}^0$ spacetime}

\begin{proposition}\label{prop 5.5}
    Let $(M,g)$ be a $\mathrm{C}^0$ globally hyperbolic spacetime, and let $U$ be a globally hyperbolic neighbourhood of $p\in M$ covered by an $\epsilon$-simple coordinate system $\{x^0, x^1, \cdots, x^n\}$ at $p$, where $\epsilon < \frac{1}{100n^2}$. Let $O_{\varphi}^{+}$ denote the coordinate future timelike cone of hyperbolic angle $\varphi$ (see Figure \ref{fig 2}), and let $\{q_k\} \subset O_{\varphi}^{+}$ be a sequence of points converging to $p$ in $U$.

    \begin{figure}[h]
        \centering
        \begin{tikzpicture}
            \draw[->] (-3,0) -- (3,0);
            \draw[->] (0,0) -- (0,3.5);
            %%%%%
            \node[below] at (0,0) {\footnotesize $o$};
            %%%%%
            \draw[domain=0:2.5,smooth,variable=\y]
            plot ({\y},{\y});
            \draw[domain=0:2.5,smooth,variable=\y]
            plot ({-\y},{\y});
            %%%%%
            \draw[domain=0:3,smooth,variable=\y]
            plot ({0.6*\y},{\y});
            \draw[domain=0:3,smooth,variable=\y]
            plot ({-0.6*\y},{\y});
            \node[left] at ({0.6*1.8},{1.8}) {\footnotesize $O_{\varphi}^+$};
            %%%%%
            \path[fill=blue,opacity=0.2]
            (0,0)
            --
            ({0.6*3},{3})
            --
            ({-0.6*3},{3});
        \end{tikzpicture}
        \caption{Coordinate future timelike cone of hyperbolic angle $\varphi$}
        \label{fig 2}
    \end{figure}
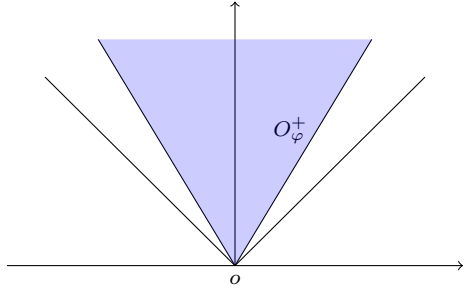

    \begin{enumerate}[label=(\alph*)]
        \item
              There exists $K$ such that for all $k \geq K$, we have $q_k \in I^{+}(p)$.

        \item
              For $k \geq K$, define
              \begin{align*}
                  d_k = d(p,q_k,U),
                  \quad
                  V_k = \vert I(p,q_k,U)\vert,
                  \quad
                  Q_k = \vert J(p,q_k,U)\vert.
              \end{align*}
              where $d(\cdot,\cdot,U)$ denotes the distance in $(U,g)$. The volumes $V_k$ and $Q_k$ exhibit the following asymptotic behavior as $k\rightarrow +\infty$:
              \begin{align*}
                  \frac{V_k}{ d_k^{n+1} },\frac{Q_k}{d_k^{n+1}}
                  =
                  \frac{2\omega_n}{n+1}  \cdot \left(\frac{1}{2}\right)^{n+1} + o (1).
              \end{align*}
    \end{enumerate}
\end{proposition}

\begin{proof}
    \begin{enumerate}[label=(\alph*)]
        \item
              Consider the compact set $O_{\varphi, v^0=1}^+$ in $O_{\varphi}^+$:
              \begin{align*}
                  O_{\varphi, v^0=1}^+
                  =
                  \{ v \in O_{\varphi}^+: v^0=1\}.
              \end{align*}
              Since
              $\sup_{v\in O_{\varphi,v^0=1}^+} \eta (v,v) <0$,
              the continuity of $g$ implies the existence of a neighbourhood $U'$ of $o$ such that
              \begin{align*}
                  \sup_{x\in U', v\in O_{\varphi,v^0=1}^+} g(x) (v,v) <0.
              \end{align*}
              There exists $K$ such that for all $k\geq K$, the line $\overline{o\phi(q_k)} \subset U'$. Define the curve $\gamma$ from $o$ to $\phi(q_k)$:
              \begin{align*}
                  \gamma:
                  [0,1] \rightarrow V',
                  \quad
                  \gamma(t) = (1-t) o + t \phi(q_k).
              \end{align*}
              Then $\gamma$ is timelike in $(U',g)$, so $q_k \in I^{+}(p)$ for all $k\geq K$.

        \item
              For any $\delta > 0$, there exists a neighbourhood $U_{\delta} \subset U$ of $p$ such that for all $x\in U_{\delta}$,
              \begin{align*}
                  \vert g_{uv}(x) - \eta_{uv} \vert
                  <
                  \delta.
              \end{align*}
              Then by the global hyperbolicity, there exists a sufficiently small neighbourhood $V_{\delta} \subset U_{\delta}$ of $p$ such that for any $q_k \in V_{\delta}$,
              \begin{align*}
                   &
                  J(p,q_k,U) = J(p,q_k,U_{\delta}),
                  \quad
                  I(p,q_k,U) = I(p,q_k,U_{\delta}),
                  \\
                   &
                  d_k = d(p,q_k,U) = d( p, q_k, U_{\delta}).
              \end{align*}
              Let $O^+_{\varphi, \vert v \vert_{\eta}=1}$ denote the set of timelike directions in $O^+_{\varphi}$:
              \begin{align*}
                  O^+_{\varphi, \vert v \vert_{\eta}=1}
                  =
                  \{ v \in O^+_{\varphi}:  \vert v \vert_{\eta} = 1 \}.
              \end{align*}
              Denote the Minkowski length of $\overline{o \phi(q_k)}$ by
              $d_k' = \vert \overline{o \phi(q_k)} \vert_{\eta}$.
              By the compactness of $O^+_{\varphi, \vert v \vert_{\eta}=1}$ and Lemmas \ref{lem 5.3} and \ref{lem 5.4}, there exist constants $C$ and $\delta_0$ such that if $\delta < \delta_0$ and $q_k \in V_{\delta}$,
              \begin{align*}
                   &
                  \vert d_k - d_k' \vert
                  \leq
                  C \delta d_k,
                  \\
                   &
                  \big\vert
                  V_k
                  -
                  \vert I_{\eta}(o,q_k) \vert_{\eta}
                  \big\vert
                  \leq
                  C \delta {d_k'}^{n+1},
                  \\
                   &
                  \big\vert
                  Q_k
                  -
                  \vert J_{\eta}(o,q_k) \vert_{\eta}
                  \big\vert
                  \leq
                  C \delta {d_k'}^{n+1}.
              \end{align*}
              Consequently,
              \begin{align*}
                  V_k
                  =
                  \frac{2\omega_n}{n+1} \left(\frac{d_k'}{2}\right)^{n+1}
                  +  O(\delta {d_k'}^{n+1})
                  =
                  \frac{2\omega_n}{n+1} \left(\frac{d_k}{2}\right)^{n+1}
                  +  O(\delta {d_k}^{n+1}).
              \end{align*}
              Taking $\delta \to 0$ as $d_k \to 0$ yields the asymptotic behavior of $\frac{V_k}{d_k^{n+1}}$. Similar for the asymptotic behavior of $\frac{Q_k}{d_k^{n+1}}$.
    \end{enumerate}
\end{proof}

%%%%%%%%%%%%%%%%%%%%%
%%%%%%%%%%%%%%%%%%%%%
\subsection{VDR asymptote in $\mathrm{C}^{0,1}$ spacetime}

\begin{proposition}\label{prop 5.6}
    Let $(M,g)$ be a $\mathrm{C}^{0,1}$ globally hyperbolic spacetime, and let $\gamma$ be a past-directed chronological curve with $p = \gamma(0)$. Consider a sequence $t_k$ decreasing to $0$ and define $q_k = \gamma(t_k)$, with
    \begin{align*}
        l_k
        =
        L[\gamma|_{(0,t_k)}],
        \quad
        V_k
        =
        \vert I(p,q_k) \vert.
    \end{align*}
    Then, as $k \rightarrow +\infty$, the asymptotic behavior of $V_k$ is given by
    \begin{align*}
        \frac{V_k}{ l_k^{n+1} }
        =
        \frac{2\omega_n}{n+1} \cdot \left(\frac{1}{2}\right)^{n+1} + o (1),
    \end{align*}
    provided one of the following conditions holds:
    \begin{enumerate}[label=(\alph*)]
        \item $\gamma$ is maximal.
        \item The restriction $\gamma|_{(0,\epsilon]}$ is a geodesic satisfying the geodesic equation
              $\ddot{\gamma}^{\kappa}
                  + \Gamma_{\mu\nu}^{\kappa} \dot{\gamma}^{\mu} \dot{\gamma}^{\nu}
                  =
                  0$,
              where $\Gamma_{\mu\nu}^{\kappa}$ and and the differentials $\ed g_{\mu\nu}$ exist along $\gamma|_{(0,\epsilon]}$.
    \end{enumerate}
\end{proposition}

\begin{proof}
    Let $d_k = d(p,q_k)$.
    \begin{enumerate}[label=(\alph*)]
        \item
              If $\gamma$ is maximal, then $d_k = l_k$. By Proposition \ref{prop 5.5}, it suffices to show that there exists a simple coordinate system at $p$ and a coordinate future timelike cone $O^+_{\phi}$ such that $q_k \in O^+_{\phi}$ for all sufficiently large $k$.

              Choose a simple coordinate system at $p$. By Theorem 1.1 of \cite{LLS21}, $\gamma$ admits a $\mathrm{C}^{1,1}$ reparameterization with constant velocity. Without loss of generality, take $t$ to be the arc-length parameter so that $\vert \gamma' \vert = 1$. Then, by the continuity and timelike nature of $\gamma'$, there exists a coordinate future timelike cone $O^+_{\phi}$ containing $q_k$ for sufficiently large $k$.

        \item
              By Lemma \ref{lem 4.25}, $\gamma$ is Lipschitz-continuously differentiable on $[0,\epsilon]$, and $\dot{\gamma}(0)$ is future timelike. Choose a simple coordinate system at $p$ such that $\dot{\gamma}(0) = \partial_0$. Then
              $\gamma^{\kappa}(t) = \delta_0^{\kappa} t + O(t^2)$,
              $L[\gamma|_{[0,t]}] = t + O(t^2)$.
              By an argument similar to that in Lemma \ref{lem 5.3}, we obtain
              $d(p,\gamma(t)) = t + O(t^2)$,
              and hence
              $\lim_{k\rightarrow +\infty}\frac{d_k}{l_k} = 1$.
              The asymptotic result then follows from Proposition \ref{prop 5.5}.
    \end{enumerate}
\end{proof}

%%%%%%%%%%%%%%%%%%%%%
%%%%%%%%%%%%%%%%%%%%%
%%%%%%%%%%%%%%%%%%%%%
\section{Criteria of strongly-causal inextendibility}

%%%%%%%%%%%%%%%%%%%%%
%%%%%%%%%%%%%%%%%%%%%
\subsection{Criterion of $\mathrm{C}^0$ locally null-non-accumulating strongly-causal inextendibility}

\begin{theorem}\label{thm 6.1}
    Let $(M,g)$ be a $\mathrm{C}^2$ globally hyperbolic spacetime with a future boundary point $\hat{q}$. There exists no local $\mathrm{C}^0$ locally null-non-accumulating strongly-causal extension of $I^-(\hat{q})$ at $\hat{q}$ if for every future-ordering chronological sequence $\{ q_k \}$ exhausting $\hat{q}$, we have
    \begin{align*}
        \frac{V_k}{ d_k^{n+1} } \nrightarrow \frac{2\omega_n}{n+1} \cdot (\frac{1}{2})^{n+1},
    \end{align*}
    where
    $V_k = \vert I(q_k, \hat{q}) \vert$,
    and
    $d_k = d(q_k, \hat{q})$.
\end{theorem}

\begin{proof}
    Suppose, for contradiction, that $(\tilde{M},\tilde{g},\phi)$ is a local $\mathrm{C}^0$ locally null-non-accumulating strongly-causal extension of $I^-(\hat{q})$ at $\hat{q}$. Denote $\phi(\hat{q})$ by $\tilde{q}$. By Proposition \ref{prop 3.10}, $\phi$ is FBC at $\hat{q}$.

    Propositions \ref{prop 4.17} and \ref{prop 4.20} imply that $\phi$ is chronological-diamond-preserving and locally chronological-diamond-surjective at $\hat{q}$. Choose a globally hyperbolic chronologically convex neighbourhood $\tilde{U}$ of $\tilde{q}$, covered by an $\epsilon$-simple ($\epsilon < \frac{1}{100n^2}$) coordinate system at $\tilde{q}$ of $(\tilde{M}, \tilde{g})$, such that for any $p\ll q \in \phi^{-1}(\tilde{U})$,
    \begin{align*}
        I(\phi(p),\phi(q),\tilde{U}) = \phi(I(p,q,M)),
        \quad
        I(\phi(p),\tilde{q},\tilde{U}) \subset \phi(I(p,\hat{q})) \subset J(\phi(p),\tilde{q}, \tilde{U}).
    \end{align*}
    The existence of $\tilde{U}$ follows from Propositions \ref{prop 3.12} and \ref{prop 4.16}.

    There exists a future-ordering chronological sequence $\{q_k\}\subset \phi^{-1}(\tilde{U})$ exhausting $\hat{q}$, %in the chronological past, 
    and a future-ordering chronological sequence $\{\tilde{p}_k\} \subset I^-(\tilde{q},\tilde{U})$, such that
    $I(\tilde{p}_k, \tilde{q}, \tilde{U})
        \subset
        \phi(I(q_k,\hat{q}))$.
    Let $\gamma$ be the coordinate line
    $\gamma (t) = (t,0,\cdots,0)$.
    For sufficiently small $\epsilon$, $\gamma|_{[-\epsilon,0]}$ is chronological. Moreover, there exists an increasing sequence $\{t_k>-\epsilon\}$ converging to $0$ such that
    $\gamma|_{[t_k,0)}
        \subset
        I(\tilde{p}_k, \tilde{q}, \tilde{U})
        \subset
        \phi(I(q_k,\hat{q}))$.
    Define the future-ordering chronological sequence $\{ q'_k \}$ by
    $q'_k = \phi^{-1} (\gamma(t_k))$.
    Then $\{q'_k\}$ exhausts $\hat{q}$.
    Define
    \begin{align*}
         &
        d_{k,l} = d(q'_k,q'_l, M),
        \quad
        \tilde{d}_{k,l} = d(\phi(q'_k),\phi(q'_l), \tilde{U}),
        \quad
        \tilde{d}_k = d(\phi(q'_k),\tilde{q}, \tilde{U}),
        \\
         &
        \tilde{V}_k = I(\phi(q'_k),\tilde{q},\tilde{U}),
        \quad
        \tilde{Q}_k = J(\phi(q'_k),\tilde{q},\tilde{U}).
    \end{align*}
    We have
    $d_{k,l} = \tilde{d}_{k,l}$,
    $\tilde{V}_k \leq V_k \leq \tilde{Q}_k$.
    Since $\{q'_k\}$ exhausts $\hat{q}$, $\{ \phi(q'_k) = \gamma(t_k) \}$ converges to $\tilde{q}$ from the chronological past in $\tilde{U}$, we conclude that
    \begin{align*}
        d_k = \lim_{l\rightarrow \infty} d_{k,l},
        \quad
        \tilde{d}_k = \lim_{l\rightarrow \infty} \tilde{d}_{k,l}.
    \end{align*}
    Hence
    $d_k = \tilde{d}_k$.
    Therefore, by Proposition \ref{prop 4.5},
    $\frac{V_k}{d_k^{n+1}} \rightarrow \frac{2\omega_n}{n+1} \cdot (\frac{1}{2})^{n+1}$,
    contradicting the assumption.
\end{proof}

%%%%%%%%%%%%%%%%%%%%%
%%%%%%%%%%%%%%%%%%%%%
\subsection{Criterion of $\mathrm{C}^{0,1}$ strongly-causal inextendibility}

\begin{theorem}\label{thm 6.2}
    Let $(M,g)$ be a $\mathrm{C}^2$ globally hyperbolic spacetime with a future boundary point $\hat{q}$. No local $\mathrm{C}^{0,1}$ strongly-causal extension of $I^-(\hat{q})$ exists at $\hat{q}$ if there is a future-directed chronological geodesic curve $\gamma: [t_0,0) \rightarrow M$ exhausting $\hat{q}$ such that as $t\rightarrow 0^-$,
    \begin{align*}
        \frac{V(t)}{ l(t)^{n+1} } \nrightarrow \frac{2\omega_n}{n+1} \cdot (\frac{1}{2})^{n+1},
    \end{align*}
    where
    $V(t) = \vert I(\gamma(t), \hat{q}) \vert$,
    $l(t) = L[\gamma|_{[t,0)}]$.
\end{theorem}

\begin{proof}
    Assume by contradiction that $(\tilde{M},\tilde{g},\phi)$ is a local $\mathrm{C}^{0,1}$ strongly-causal extension of $I^-(\hat{q})$ at $\hat{q}$. Let $\tilde{q} = \phi(\hat{q})$. By Proposition \ref{prop 4.26}, $\phi$ is null-non-accumulating, and hence chronological-diamond-preserving at $\hat{q}$.

    Defin
    $\tilde{\gamma} = \phi \circ \gamma$.
    By Lemma \ref{lem 4.25}, $\tilde{\gamma}$ is Lipschitz-continuously differentiable at $t=0$ with future timelike tangent vector $\dot{\tilde{\gamma}}(0)$.
    Following the same argument as in the proof of Theorem \ref{thm 6.1} and Proposition \ref{prop 5.6}, we obtain
    $\frac{V(t)}{l(t)^{n+1}}
        \rightarrow
        \frac{2\omega_n}{n+1} \cdot (\frac{1}{2})^{n+1}$,
    which contradicts the assumption.
\end{proof}

%%%%%%%%%%%%%%%%%%%%%
%%%%%%%%%%%%%%%%%%%%%
\subsection{Application: $\mathrm{C}^0$ strongly-causal inextendibility of Misner spacetime}

\begin{theorem}\label{thm 6.3}
    The Misner spacetime in Example \ref{ex 4.22} admits no $\mathrm{C}^0$ strongly-causal extension at the future boundary point $o$.
\end{theorem}

\begin{proof}
    We first compute the VDR asymptote of the Misner spacetime $(M,\eta)$ at $o$ along any future-directed chronological sequence. Due to the scale invariance and Lorentz rotational symmetry of $(M,\eta)$, it suffices to compute
    $\frac{\vert I(p,o) \vert}{\vert d(p,o) \vert^2}$
    at a single point $p$, as the VDR of $I(p,o)$ is identical for all points.

    Moreover, since $(M,g)$ arises as the quotient of the $2$-dimensional Minkowski spacetime, the VDR of $I(p,o)$ is strictly smaller than the Minkowski value:
    $\frac{\vert I(p,o) \vert}{\vert d(p,o) \vert^2}
        <
        \frac{1}{2}$.
    Thus, by Theorem \ref{thm 6.1}, $(M,g)$ admits no $\mathrm{C}^0$ locally null-non-accumulating strongly-causal extension at $o$.

    Since the future horismos of any $p \in (M,g)$ is compact, Proposition \ref{prop 4.21} implies that any $\mathrm{C}^0$ strongly-causal extension of $(M,g)$ at $o$ is locally null-non-accumulating. Therefore, $(M,g)$ admits no $\mathrm{C}^0$ strongly-causal extension at $o$.
\end{proof}

\begin{remark}\label{rem 6.4}
    Due to the $2$-dimensional nature of the Misner spacetime, its inextendibility can be improved to $\mathrm{C}^0$ past-distinguishing inextendibility. However, for consistency with the paper, we omit this discussion here.

    Furthermore, the same approach applies to study the $\mathrm{C}^0$ strongly-causal inextendibility of the spatially spherical FLRW spacetime in Example \ref{ex 4.23}. We defer this analysis to a subsequent paper.
\end{remark}

%%%%%%%%%%%%%%%%%%%%%
%%%%%%%%%%%%%%%%%%%%%
%%%%%%%%%%%%%%%%%%%%%
\section{Spatially flat FLRW spacetime with scale function $a(t) \sim \vert t\vert$}

This section examines a class of spatially flat FLRW spacetimes $(M,g)$, where $M = \mathbb{R}^{n+1}_{t<0}$ and
\footnote{To maintain consistency with the discussion of inextendibility at a future boundary point, we position the singularity in the future. By reversing the orientation, the results in this section readily apply to big bang singularities in the past.}
\begin{align*}
     &
    g
    =
    - \ed t^2 + a^2(t) [ (\ed x^1)^2 + \cdots + (\ed x^n)^2]
    =
    - \ed t^2 + a^2(t) ( \ed r^2 + r^2 \ringg ).
\end{align*}
where $a(t) \sim |t|$ and $\ringg$ denotes the standard round metric on the unit sphere $\mathbb{S}^{n-1}$. Since $\int_{t_0}^0  \frac{1}{a(t)} \ed t = -\infty$, for any future-directed chronological curve $\gamma(t) = (t, x^1, \cdots , x^n)$ orthogonal to the coordinate hyperplane $\Sigma_t$, we obtain
\begin{align*}
    \cap_{t<0} I^-(\gamma(t)) = M.
\end{align*}

%%%%%%%%%%%%%%%%%%%%
%%%%%%%%%%%%%%%%%%%%%
\subsection{Conformal compactification and future boundary}

Define
$\tilde{t}
    =
    A(t)
    =
    \int_{-1}^t
    \frac{1}{a(t)} \ed t$.
This yields
$\ed \tilde{t}
    =
    \frac{1}{a(t)} \ed t$.
In the coordinate system $(\tilde{t}, r, \vartheta \in \mathbb{S}^{n-1} )$,
\begin{align*}
    g
    =
    a^2(t) ( - \ed \tilde{t}^2 + \ed r^2 )
    +
    a^2 (t) r^2 \ringg
    =
    a^2(t) ( - \ed \tilde{t}^2 + \ed r^2 + r^2 \ringg).
\end{align*}
Note that $- \ed \tilde{t}^2 + \ed r^2 + r^2 \ringg$ is the Minkowski metric $\eta$, hence $g$ is conformal to $\eta$,
$g = a^2(t) \cdot \eta$.
Introduce double null coordinates $u = \frac{\tilde{t} - r}{2}$, $v= \frac{\tilde{t} + r}{2}$.
In the double null coordinate system $(u,v, \vartheta)$,
\begin{align*}
    g
    =
    - 4 a^2(t) \ed u \ed v
    +  a^2(t) r^2 \ringg
    =
    a^2(t) \cdot
    [ - 4 \ed u \ed v
        +  (v-u)^2 \ringg]
    =
    a^2(t) \cdot \eta.
\end{align*}
Consider the transformation $u = \tan \tilde{u}$, $v = \tan \tilde{v}$ where $\tilde{u}, \tilde{v} \in (-\frac{\pi}{2} ,\frac{\pi}{2} )$
which gives  $v - u
    =
    \frac{\sin (\tilde{u} - \tilde{v})}{\cos \tilde{u} \cdot \cos \tilde{v} }$ and
\begin{align*}
    g
    =
    \frac{a^2(t)}{\cos^2 \tilde{u} \cos^2 \tilde{v}}
    [ - 4 \ed \tilde{u} \ed \tilde{v}
        + \sin^2 (\tilde{u} - \tilde{v}) \ringg ].
\end{align*}

From this conformal compactification, the future boundary of $(M,g)$ consists of two components: the future timelike boundary point $\mathcal{O}$ and the future null boundary $\mathcal{FN}$.
\begin{enumerate}[label=(\alph*)]
    \item
          $\mathcal{O}$: $\tilde{u} = \tilde{v} = \frac{\pi}{2}$.

    \item
          $\mathcal{FN}$: $\tilde{u} \in (-\frac{\pi}{2}, \frac{\pi}{2} )$, $\tilde{v} = \frac{\pi}{2}$, $\vartheta \in \mathbb{S}^{n-1}$.

\end{enumerate}

%%%%%%%%%%%%%%%%%%%%
%%%%%%%%%%%%%%%%%%%%%
\subsection{$\mathrm{C}^0$ locally null-non-accumulating strongly-causal inextendibility}

We establish the $\mathrm{C}^0$ locally null-non-accumulating strongly-causal inextendibility by analyzing the VDR asymptote for small chronological diamonds $I(p,\mathcal{O})$ as $p$ approaches the future timelike boundary point $\mathcal{O}$.

%%%%%%%%%%%%%%%%%%%%
\subsubsection{Special case: $a(t) = \vert t\vert$}

We first consider the special case $a(t)=\vert t\vert$, where explicit calculations illustrate the general case. Since the metric
\begin{align*}
    g_s = - \ed t^2 + t^2 ( \ed r^2 + r^2 \ringg )
\end{align*}
is scale-invariant and spatially homogeneous, it suffices to compute the VDR asymptote for a small chronological diamond $I(p,\mathcal{O})$ at a single point $p$. The key idea is to compare our calculation with that of the spatially hyperbolic $K=-1$ FLRW spacetime, which corresponds to Minkowski spacetime:
\begin{align*}
    g_{h} = - \ed t^2 + t^2 ( \ed r^2 + \sinh^2 r \ringg ).
\end{align*}
$g_h$ is the Minkowski metric under the transformation $- t^2 = - \tilde{t}^2 + \tilde{r}^2$, $\sinh r = \tilde{r}$.

We choose
$p=(t=-1,x^1=0,\cdots x^n=0)$.
The distances from $p$ to the future timelike boundary point $\mathcal{O}$ in both $(M,g_s)$ and $(M,g_h)$ are $1$:
$d_s(p,\mathcal{O}) = d_h(p,\mathcal{O})= 1$.
The chronological diamonds $I_s(p,\mathcal{O})$ and $I_h(p,\mathcal{O})$ coincide and are denoted by $I(p,\mathcal{O})$:
\begin{align*}
    I(p,\mathcal{O})
    =
    I^+(p) \cup I_s^-(\mathcal{O})
    =
    I^+(p)
    =
    \{ r < \int_{-1}^t \frac{1}{\vert t'\vert} \ed t' = - \log |t| \}.
\end{align*}
The volume of $I(p,\mathcal{O}) \cap \Sigma_t$ for each metric is given by
\begin{align*}
    \vert I^+(p) \cap \Sigma_t \vert_{g_s}
     &
    =
    |t|^n \int_0^{- \log |t|} n \omega_n r^{n-1} \ed r
    =
    |t|^n \omega_n (- \log |t|)^n
    \\
     &
    <
    |t|^n \int_0^{- \log |t|} n \omega_n \sinh^{n-1} r \ed r
    =
    \vert I^+(p) \cap \Sigma_t \vert_{g_h}.
\end{align*}
Consequently, the volume of the chronological diamond $I(p,\mathcal{O})$ in each metric is
\begin{align*}
    \vert I(p,\mathcal{O}) \vert_{g_s}
     & =
    \int_{-1}^0 |t|^n \omega_n (- \log |t|)^n \ed t
    \\
     & <
    \int_{-1}^0 |t|^n \int_0^{- \log |t|} n \omega_n \sinh^{n-1} r \ed r \ed t
    =
    \vert I(p,\mathcal{O}) \vert_{g_h}.
\end{align*}
Since $g_h$ is the Minkowski metric,
$\vert I(p,\mathcal{O}) \vert_{g_h}
    =
    \frac{2\omega_n}{n+1} \cdot (\frac{1}{2})^{n+1}$.
Thus there exists a positive constant $\delta>0$ such that
\begin{align*}
     &
    \vert I(p,\mathcal{O}) \vert_{g_s}
    <
    \vert I(p,\mathcal{O}) \vert_{g_h} - \delta
    \quad
    \Rightarrow
    \quad
    \frac{\vert I(p,\mathcal{O}) \vert_{g_s}}{ d_s(p,\mathcal{O})^{n+1} }
    <
    \frac{2\omega_n}{n+1} \cdot (\frac{1}{2})^{n+1} - \delta.
\end{align*}
Therefore, by Theorem \ref{thm 6.1}, we conclude that $(M,g_s)$ is $\mathrm{C}^0$ null-non-accumulating strongly-causal inextendible at $\mathcal{O}$.

%%%%%%%%%%%%%%%%%%%%
\subsubsection{General case: $a(t) \sim \vert t\vert$}
We now extend the previous calculations to the general case.
\begin{theorem}\label{thm 7.1}
    The spatially flat FLRW spacetime $(M = \mathbb{R}^{n+1}_{t<0},g)$ with $n\geq 2$ and
    \begin{align*}
        g
        =
        - \ed t^2 + a^2(t) ( \ed r^2 + r^2 \ringg ),
        \quad
        a(t) \sim |t|.
    \end{align*}
    admits no local $\mathrm{C}^0$ locally null-non-accumulating strongly-causal extension at the future timelike boundary point $\mathcal{O}$.
\end{theorem}
\begin{proof}
    Let $\{p_k\}$ be a future-oriented chronological sequence exhausting $\mathcal{O}$. By the homogeneity of $g$, we may assume that
    $p_k = (t_k, x^1=0, \cdots, x^n=0)$.
    Then
    $d(p_k, \mathcal{O}) = \vert t_k \vert $.
    Define the radius $r_k(t)$ of the chronological future $I^+(p_k)$ by
    $I^+(p_k)
        =
        \{ (t,r): r< r_k(t) \}$.
    We have
    \begin{align*}
        \frac{\ed}{\ed t} r_k(t) = \frac{1}{a(t)},
        \quad
        r_k(t) = \int_{t_k}^t \frac{1}{a(t)} \ed t.
    \end{align*}
    Since $a(t) = -t + o( |t| )$, it follows that
    \begin{align*}
        r_k(t)
        =
        \int_{t_k}^t \frac{1}{a(t)} \ed t
        =
        - (1+ c(t_k, t)) \log \frac{|t|}{|t_k|}
    \end{align*}
    where $c(t_k,t) = o(1)$ as $|t_k| \rightarrow 0$. Here, $c(t',t)$ and $c(t')$ denote any functions satisfying $c(t',t) = o(1)$ and $c(t') = o(1)$ as $|t'| \rightarrow 0$. The volume of $I^+(p_k) \cap \Sigma_t$ is then
    \begin{align*}
        \vert I^+(p_k) \cap \Sigma_t \vert
         &
        =
        a^n (t) \int_0^{r_k(t)} n\omega_n r^{n-1} \ed r
        =
        a^n (t) \omega_n [r_k(t)]^n
        \\
         &
        =
        ( 1 + c(t_k, t) ) |t|^n \cdot \omega_n \big(-\log \frac{|t|}{|t_k|} \big)^n.
    \end{align*}
    Thus, the volume of the chronological diamond $I(p_k,\mathcal{O})$ is
    \begin{align*}
        \vert I(p_k,\mathcal{O}) \vert
         &
        =
        \vert I^+(p_k) \vert
        =
        \int_{t_k}^0
        ( 1 + c(t_k,t) ) |t|^n \cdot \omega_n \big(-\log \frac{|t|}{|t_k|}\big)^n \ed t
        \\
         &
        =
        ( 1 + c(t_k) )  |t_k|^{n+1} \int_{-1}^0 |t'|^n \cdot \omega_n (-\log |t'|)^n \ed t'.
    \end{align*}
    Consequently, the VDR asymptote of $I(p_k,\mathcal{O})$ satisfies
    \begin{align*}
        \lim_{t_k \rightarrow0^-}
        \frac{\vert I(p_k,\mathcal{O}) \vert}{ [d(p_k,\mathcal{O})]^{n+1} }
        =
        \int_{-1}^0 |t'|^n \cdot \omega_n (-\log |t'|)^n \ed t'
        <
        \frac{2\omega_n}{n+1} \cdot (\frac{1}{2})^{n+1} - \delta.
    \end{align*}
    Therefore, by Theorem \ref{thm 6.1}, the result follows.
\end{proof}

%%%%%%%%%%%%%%%%%%%%%
%%%%%%%%%%%%%%%%%%%%%
%%%%%%%%%%%%%%%%%%%%%
\section{Self-similar naked singularity in the gravitational collapse of a scalar field}

Recall Christodoulou's example of self-similar naked singularities arising from gravitational collapse of a scalar field. In Bondi coordinates $\{u\leq 0, r\geq 0, \vartheta \in \mathbb{S}^2\}$
the spherically symmetric metric is given by
\begin{align*}
    \ed s^2= -e^{2\nu} \ed u^2 - 2e^{\nu+\lambda} \ed u \ed r + r^2 \ringg,
\end{align*}
where $\ringg$ denotes the standard round metric on the unit sphere $\mathbb{S}^2$. Let $\mathcal{Q}$ be the two-dimensional quotient spacetime with Lorentzian metric $g$ on $\mathcal{Q}$:
$g =  -e^{2\nu} \ed u^2 - 2e^{\nu+\lambda} \ed u \ed r$.
In the self-similar coordinate system $\{u, x, \vartheta\}$, where $x = -\frac{r}{u}$, $r = -ux$,
the metric becomes
\begin{align*}
     &
    \ed s^2
    =
    e^{\nu + \lambda}
    ( 2x \beta \ed u^2 + 2 u \ed u \ed x )
    +
    (ux)^2 \ringg,
    \\
     &
    \nu = \nu(x),
    \quad
    \lambda = \lambda(x),
    \quad
    \beta
    =
    1
    -
    \frac{e^{\nu - \lambda}}{2x}.
\end{align*}
The volume form $\dvol$ of the metric $\ed s^2$ is
\begin{align*}
     &
    \dvol
    =
    2 e^{\nu+\lambda} x^2 u^3 \ed u \ed x \ed \sigma^2
    =
    4 e^{2\lambda} \cdot (1-\beta) \cdot x^3 u^3 \ed u \ed x \ed \sigma^2,
\end{align*}
with $\ed \sigma^2$ being the volume form of $(\mathbb{S}^2, \ringg)$.

In the self-similar coordinate system, the Minkowski metric takes the form
\begin{align*}
     &
    \nu = \lambda = 0,
    \quad
    \beta = 1 - \frac{1}{2x},
    \\
     &
    \eta
    =
    2x \beta \ed u^2
    +
    2u \ed u \ed x
    +
    (ux)^2 \ringg
    =
    ( 2x - 1 ) \ed u^2
    +
    2u \ed u \ed x
    +
    (ux)^2 \ringg,
\end{align*}
and the corresponding Minkowski volume form $\dvol_{\eta}$ is
\begin{align*}
     &
    \dvol_{\eta}
    =
    2 x^2 u^3 \ed u \ed x \ed \sigma^2
    =
    4 (1-\beta) x^3 u^3 \ed u \ed x \ed \sigma^2.
\end{align*}

%%%%%%%%%%%%%%%%%%%%
%%%%%%%%%%%%%%%%%%%%%
\subsection{Reduction of self-similar Einstein-scalar field equations}

Following Christodoulou, we introduce the variables $\chi$, $\theta$, and $s$ to derive an autonomous ODE system equivalent to the Einstein equations coupled to a massless scalar field:
\begin{align*}
    \phi = \chi(x) - k \log (-u),
    \quad
    \theta = x \frac{\ed \chi}{\ed x},
    \quad
    x = e^s.
\end{align*}
This yields the autonomous system for $\{\beta, \theta\}$:
\begin{align*}
    \left\{
    \begin{aligned}
         &
        \frac{\ed \beta}{\ed s}
        =
        x \frac{\ed \beta}{\ed x}
        =
        1 - k^2 - [ (\theta + k)^2 + 1 -k^2 ] \beta,
        \\
         &
        \frac{\ed \theta}{\ed s}
        =
        x \frac{\ed \theta}{\ed x}
        =
        k \beta^{-1} (k \theta -1) + \theta [(\theta+k)^2 - (1+k^2)].
    \end{aligned}
    \right.
\end{align*}
Introducing the variable
$\alpha = \beta^{-1}$,
we rewrite the system in terms of $\{\alpha, \theta\}$:
\begin{align*}
    \left\{
    \begin{aligned}
         &
        \frac{\ed \alpha}{\ed s}
        =
        x \frac{\ed \alpha}{\ed x}
        =
        \alpha [(\theta+k)^2 + (1-k^2) (1-\alpha)],
        \\
         &
        \frac{\ed \theta}{\ed s}
        =
        x \frac{\ed \theta}{\ed x}
        =
        k \alpha (k \theta -1) + \theta [(\theta+k)^2 - (1+k^2)].
    \end{aligned}
    \right.
\end{align*}
Under the boundary conditions
$\lim_{s \rightarrow - \infty} (\alpha e^{-s} ) = -2$,
$\lim_{s\rightarrow - \infty} \theta = 0$,
the behavior of solutions splits into two cases:
\begin{enumerate}[label=(\alph*)]
    \item
          For $0< k <1$, the solution approaches the singular point $(\beta = 0, \theta = \frac{1}{k})$ at a finite value $s_*$. The interval $(-\infty, s_*)$ corresponds to the chronological past of the scaling origin $\mathcal{O} = (u=0,r=0)$, with the singular point forming the future boundary of the interior solution. Within this interval, $\beta$ increases monotonically from $-\infty$ to $0$, while $\theta$ remains positive and satisfies $\lim_{s\rightarrow s_*} \theta = \frac{1}{k}$.

    \item
          For $k \geq 1$, the solution is defined for all $s \in \mathbb{R}$ with $\beta<0$ everywhere. In this case, $\theta$ is bounded and positive for all $s$. The scaling origin $\mathcal{O}$ lies outside the domain of dependence of the interior of any sphere $(u=u_0, x=x_0)$ on the cone $u=u_0$.

\end{enumerate}
The following identity will be useful later:
\begin{align*}
     &
    e^{2\lambda} - 1
    =
    k^2 + \frac{\beta}{1-\beta} (\theta + k)^2
    =
    k^2 + \frac{1}{\alpha -1} (\theta + k)^2.
\end{align*}

%%%%%%%%%%%%%%%%%%%%
%%%%%%%%%%%%%%%%%%%%%
\subsection{Taylor expansions of $\alpha$ and $\theta$ at the axis}

\begin{lemma}[Taylor expansions of $\alpha$ and $\theta$ up to $x^2$]\label{lem 8.1}
    \begin{align*}
         &
        \alpha
        =
        \alpha_{(1)} x
        +
        \alpha_{(2)} x^2
        +
        O(x^3),
        \quad
        \theta
        =
        \theta_{(1)} x
        +
        \theta_{(2)} x^2
        +
        O(x^3),
        \\
         &
        \alpha_{(1)}
        =
        -2,
        \quad
        \alpha_{(2)}
        =
        -4,
        %\\
        % &
        \quad
        \theta_{(1)}
        =
        k,
        \quad
        \theta_{(2)}
        =
        \frac{4}{3}k.
    \end{align*}
\end{lemma}
\begin{proof}
    The asymptotic behavior of $\alpha$ at $x=0$, given by $\lim_{s \rightarrow -\infty} (\alpha e^{-s}) = -2$, implies that $\alpha \sim -2x$.
    Hence, $\alpha_{(1)} = -2$. Substituting the expansions of $\alpha$ and $\theta$ into the system yields
    \begin{align*}
         &
        \alpha_{(1)} x
        +
        2 \alpha_{(2)} x^2
        =
        \alpha_{(1)} x
        +
        \{ \alpha_{(2)} + \alpha_{(1)} [ 2 k \theta_{(1)} - (1-k^2) \alpha_{(1)} ]  \} x^2
        +
        O(x^3),
        \\
         &
        \theta_{(1)} x
        +
        2 \theta_{(2)} x^2
        =
        [- k \alpha_{(1)}
        -
        \theta_{(1)} ] x
        +
        [- k \alpha_{(2)}
        -
        \theta_{(2)}
        +
        k^2 \alpha_{(1)} \theta_{(1)}
        +
        2 k  \theta_{(1)}^2 ] x^2
        +
        O(x^3),
    \end{align*}
    It follows that
    \begin{align*}
         &
        2 \alpha_{(2)}
        =
        \alpha_{(2)}
        +
        \alpha_{(1)} [ 2 k \theta_{(1)} - (1-k^2) \alpha_{(1)} ],
        \\
         &
        \theta_{(1)}
        =
        -k \alpha_{(1)} - \theta_{(1)},
        \quad
        2 \theta_{(2)}
        =
        -
        k \alpha_{(2)}
        -
        \theta_{(2)}
        +
        k^2 \alpha_{(1)} \theta_{(1)}
        +
        2 k  \theta_{(1)}^2.
    \end{align*}
    Solving these gives $\alpha_{(2)}$, $\theta_{(1)}$, $\theta_{(2)}$.
\end{proof}

%%%%%%%%%%%%%%%%%%%%
%%%%%%%%%%%%%%%%%%%%%
\subsection{Volume form estimate}

We estimate the volume form by comparing it with the Minkowski volume form in the coordinate system $\{ u, \beta, \vartheta \}$. The Minkowski volume form is given by
\begin{align*}
    \dvol_{\eta}
    =
    \frac{1}{4} \frac{u^3}{(1-\beta)^4} \ed u \ed \beta \ed \sigma^2.
\end{align*}
We compute the volume form $\dvol$ in $\{ u, \beta, \vartheta \}$ coordinates:
\begin{align*}
     &
    \ed x = \frac{x}{1-\beta} \cdot \frac{1}{2-e^{2\lambda}} \ed \beta,
    \quad
    \dvol
    =
    4 u^3 x^4 \cdot \frac{e^{2\lambda}}{2-e^{2\lambda}} \ed u \ed \beta \ed \sigma^2.
\end{align*}
We demonstrate that $\dvol > \dvol_{\eta}$ for $\beta<0$, as established by the following lemmas.

\begin{lemma}\label{lem 8.2}
    For $0< k <1$ and $x>0$ in the interior solution,
    $e^{2\lambda} > 1$,
    which is equivalent to
    $k^2 \alpha + (\theta+k)^2 - k^2 <0$.
\end{lemma}

\begin{proof}
    We have
    \begin{align*}
        e^{2\lambda} - 1
        =
        k^2 + \frac{1}{\alpha -1} (\theta + k)^2
        =
        \frac{k^2 \alpha + (\theta+k)^2 - k^2}{\alpha -1}.
    \end{align*}
    Define $f(x) = k^2 \alpha + (\theta+k)^2 - k^2$. To establish $f(x)<0$ for $x>0$ in the interior solution, we proceed as follows:

    \begin{enumerate}[label=(\alph*)]
        \item
              Taylor expansion of $f(x)$ at $x=0$:
              \begin{align*}
                  f
                   &
                  =
                  k^2 [ -2x -4x^2 + O(x^3) ]
                  +
                  [ k + kx + \frac{4}{3}k x^2 + O(x^3) ]^2
                  -
                  k^2
                  +
                  O(x^3)
                  \\
                   &
                  =
                  -
                  \frac{1}{3}k^2 x^2
                  +
                  O(x^3).
              \end{align*}

        \item
              Show that $f'(x) < 0$ when $f(x)=0$ and $x>0$ in the interior solution.
              We compute:
              \begin{align*}
                  x f'(x)
                   &
                  =
                  2(\theta + k)
                  \{ k \alpha (k \theta -1)
                  +
                  \theta [(\theta+k)^2 - (1+k^2)] \}
                  \\
                   & \phantom{=}
                  +
                  k^2 \alpha
                  [ (\theta+k)^2 + (1-k^2) (1-\alpha) ].
              \end{align*}
              Substituting $f = k^2 \alpha + (\theta+k)^2 - k^2 = 0$, we obtain:
              \begin{align*}
                  x f'(x)
                   &
                  =
                  2(\theta + k)
                  ( - k \alpha - \theta  )
                  +
                  \alpha (\theta+k)^2
                  =
                  (\theta + k)
                  [-2\theta + \alpha ( \theta - k) ]
                  \\
                   &
                  =
                  \frac{1}{k^2} (\theta + k)
                  [-2k^2 \theta - \theta ( \theta + 2k ) ( \theta - k) ]
                  =
                  -\frac{1}{k^2} \theta^2 (\theta + k)^2
                  <
                  0.
              \end{align*}
              This inequality holds for $x>0$ in the interior solution.
    \end{enumerate}
    The two points above imply that $f(x)<0$ for $x>0$ in the interior solution.
\end{proof}

\begin{lemma}\label{lem 8.3}
    For $0< k <1$ and $x>0$ in the interior solution,
    $2x
        >
        \frac{1}{1-\beta}$.
\end{lemma}
\begin{proof}
    Using the equation for $\beta$ and Lemma \ref{lem 8.2}, for $x>0$ in the interior solution:
    \begin{align*}
         &
        x \frac{\ed}{\ed x} \beta
        =
        (2 - e^{2\lambda}) (1-\beta)
        <
        1 - \beta
        \quad
        \Rightarrow
        \quad
        \frac{\ed}{\ed x} \ln [ x (1-\beta) ]
        >
        0.
    \end{align*}
    Therefore
    $\ln [ x (1-\beta) ]
        >
        \lim_{x\rightarrow 0^+} \ln [ x (1-\beta) ]
        =
        - \ln 2$,
    by $\beta = \alpha^{-1} = \frac{1}{-2 x + O(x^2)}$.
\end{proof}

We now derive the volume form estimate as a corollary of Lemmas \ref{lem 8.2} and \ref{lem 8.3}. The proof is straightforward.
\begin{proposition}\label{prop 8.4}
    For $0< k <1$ and $x>0$ in the interior solution,
    \begin{align*}
        \dvol
        =
        4 u^3 x^4 \cdot \frac{e^{2\lambda}}{2-e^{2\lambda}} \ed u \ed \beta \ed \sigma^2
        >
        \frac{1}{4} \frac{u^3}{(1-\beta)^4} \ed u \ed \beta \ed \sigma^2
        =
        \dvol_{\eta}.
    \end{align*}
\end{proposition}

%%%%%%%%%%%%%%%%%%%%
%%%%%%%%%%%%%%%%%%%%%
\subsection{$\mathrm{C}^{0,1}$ strongly-causal inextendibility at self-similar naked singularity}

\begin{theorem}\label{thm 8.5}
    For $0 < k < 1$, the interior solution admits no local $\mathrm{C}^{0,1}$ strongly-causal extension at the scaling origin $\mathcal{O}$.
\end{theorem}
\begin{proof}
    We apply Theorem \ref{thm 6.2}. The central axis
    $\gamma(t) = (u=t, x=0)$
    is a chronological geodesic parametrized by arc length that converges to the scaling origin $\mathcal{O}$ as $t \rightarrow 0^-$. Let $V(t)$ be as in Theorem \ref{thm 6.2}. By the volume form estimate in Proposition \ref{prop 8.4} and the fact that $\frac{\dvol}{\dvol_{\eta}}$ is independent of $u$, there exists a positive constant $\delta$ such that
    \begin{align*}
        V(t)
        >
        \frac{\omega_3}{2} \left(\frac{t}{2}\right)^4
        +
        \delta t^4.
    \end{align*}
    Hence, as $t\rightarrow 0^{-}$,
    $\frac{V(t)}{t^4}
        >
        \frac{\omega_3}{2^5} + \delta$,
    which shows that $\frac{V(t)}{t^4} \nrightarrow \frac{\omega_3}{2^5}$. The $\mathrm{C}^{0,1}$ strongly-causal inextendibility then follows from Theorem \ref{thm 6.2}.
\end{proof}

%%%%%%%%%%%%%%%%%%%%
%%%%%%%%%%%%%%%%%%%%
%%%%%%%%%%%%%%%%%%%%
\section*{Acknowledgements}
\addcontentsline{toc}{section}{Acknowledgements}
The author is supported by the National Key R\&D Program of China 2024YFA1015000 and the National Natural Science Foundation of China 12201338.

\Address


\begin{thebibliography}{100}
    \setlength{\itemsep}{0pt}      % Space between items
    \setlength{\parskip}{0pt}      % Space between paragraphs
    \setlength{\parsep}{0pt}       % Space within items
    \small

    \bibitem[A25]{A25}
    An, Xinliang:
    Naked singularity censoring with anisotropic apparent horizon.
    \textit{Ann. of Math.} (2) \textbf{201} (2025), no. 3, 775–908.

    \bibitem[BEE96]{BEE96}
    Beem, John K.; Ehrlich, Paul E.; Easley, Kevin L.:
    \textit{Global Lorentzian geometry}.
    Second edition
    Monogr. Textbooks Pure Appl. Math., 202
    Marcel Dekker, Inc., New York, 1997. xiv+635 pp.

    \bibitem[Cho52]{Cho52}
    Choquet-Bruhat, Yvonne:
    Th\'eor\`eme d'existence pour certains syst\`emes d'\'equations aux d\'eriv\'ees partielles non lin\'eaires.
    \textit{Acta Math.} \textbf{88} (1952), 141–226.

    \bibitem[Chr91]{Chr91}
    Christodoulou, Demetrios:
    The formation of black holes and singularities in spherically symmetric gravitational collapse.
    \textit{Comm. Pure Appl. Math.} \textbf{44} (1991), no. 3, 339–373.

    \bibitem[Chr93]{Chr93}
    Christodoulou, Demetrios:
    Bounded variation solutions of the spherically symmetric Einstein-scalar field equations.
    \textit{Comm. Pure Appl. Math.} \textbf{46} (1993), no. 8, 1131–1220.

    \bibitem[Chr94]{Chr94}
    Christodoulou, Demetrios:
    Examples of naked singularity formation in the gravitational collapse of a scalar field.
    \textit{Ann. of Math.} (2) \textbf{140} (1994), no. 3, 607–653.

    \bibitem[Chr99a]{Chr99a}
    Christodoulou, Demetrios:
    The instability of naked singularities in the gravitational collapse of a scalar field.
    \textit{Ann. of Math.} (2) \textbf{149} (1999), no. 1, 183–218.

    \bibitem[Chr99b]{Chr99b}
    Christodoulou, Demetrios:
    On the global initial value problem and the issue of singularities.
    \textit{Classical Quantum Gravity} \textbf{16} (1999), no. 12A, A23–A36.

    \bibitem[Chr09]{Chr09}
    Christodoulou, Demetrios:
    \textit{The formation of black holes in general relativity.}
    EMS Monographs in Mathematics.
    European Mathematical Society (EMS), Z\"urich, 2009. x+589 pp.

    \bibitem[CG12]{CG12}
    Chru\'sciel, Piotr T.; Grant, James D. E.:
    On Lorentzian causality with continuous metrics.
    \textit{Classical Quantum Gravity} \textbf{29} (2012), no. 14, 145001, 32 pp.

    \bibitem[D03]{D03}
    Dafermos, Mihalis:
    Stability and instability of the Cauchy horizon for the spherically symmetric Einstein-Maxwell-scalar field equations.
    \textit{Ann. of Math.} (2) \textbf{158} (2003), no. 3, 875–928.

    \bibitem[DL25]{DL25}
    Dafermos, Mihalis; Luk, Jonathan:
    The interior of dynamical vacuum black holes I: The $\mathrm{C}^0$-stability of the Kerr Cauchy horizon.
    \textit{Ann. of Math.} (2) \textbf{202} (2025), no. 2, 309–630.

    \bibitem[FS12]{FS12}
    Fathi, Albert; Siconolfi, Antonio:
    On smooth time functions.
    \textit{Math. Proc. Cambridge Philos. Soc.} \textbf{152} (2012), no. 2, 303–339.

    \bibitem[GL17]{GL17}
    Galloway, Gregory J.; Ling, Eric:
    Some remarks on the $\mathrm{C}^0$-(in)extendibility of spacetimes.
    \textit{Ann. Henri Poincar\'e} \textbf{18} (2017), no. 10, 3427–3448.

    \bibitem[GLS18]{GLS18}
    Galloway, Gregory J.; Ling, Eric; Sbierski, Jan:
    Timelike completeness as an obstruction to $\mathrm{C}^0$-extensions.
    \textit{Comm. Math. Phys.} \textbf{359} (2018), no. 3, 937–949.

    \bibitem[GKP72]{GKP72}
    Geroch, Robert P.; Kronheimer, E. H.; Penrose, Roger:
    Ideal points in space-time.
    \textit{Proc. Roy. Soc. London Ser.} A \textbf{327} (1972), 545–568.

    \bibitem[GS07]{GS07}
    Gibbons, G. W.; Solodukhin, S. N.:
    The geometry of small causal diamonds.
    \textit{Phys. Lett. B} \textbf{649} (2007), no. 4, 317–324.

    \bibitem[Gr20]{Gr20}
    Graf, Melanie:
    Singularity theorems for $\mathrm{C}^1$-Lorentzian metrics.
    \textit{Comm. Math. Phys.} \textbf{378} (2020), no. 2, 1417–1450.

    \bibitem[GGKS18]{GGKS18}
    Graf, Melanie; Grant, James D. E.; Kunzinger, Michael; Steinbauer, Roland:
    The Hawking-Penrose singularity theorem for $\mathrm{C}^{1,1}$-Lorentzian metrics.
    \textit{Comm. Math. Phys.} \textbf{360} (2018), no. 3, 1009–1042.

    \bibitem[GL18]{GL18}
    Graf, Melanie; Ling, Eric:
    Maximizers in Lipschitz spacetimes are either timelike or null.
    \textit{Classical Quantum Gravity} \textbf{35} (2018), no. 8, 087001, 6 pp.

    \bibitem[GKS19]{GKS19}
    Grant, James D. E.; Kunzinger, Michael; S\"amann, Clemens:
    Inextendibility of spacetimes and Lorentzian length spaces.
    \textit{Ann. Global Anal. Geom.} \textbf{55} (2019), no. 1, 133–148.

    \bibitem[GKSS20]{GKSS20}
    Grant, James D. E.; Kunzinger, Michael; Sämann, Clemens; Steinbauer, Roland:
    The future is not always open.
    \textit{Lett. Math. Phys.}, \textbf{110} (2020), no. 1, 83–103.

    \bibitem[Gu26]{Gu26}
    Gurriaran, Sebastian:
    Non-linear instability of the Kerr Cauchy horizon near $i_+$.
    arXiv:2603.17911v2 [gr-qc] (2026).

    \bibitem[J17]{J17}
    Jubb, Ian:
    The geometry of small causal cones.
    \textit{Classical Quantum Gravity} \textbf{34} (2017), no. 9, 094005, 20 pp.

    \bibitem[KR10]{KR10}
    Klainerman, Sergiu; Rodnianski, Igor:
    On the breakdown criterion in general relativity.
    \textit{J. Amer. Math. Soc.} \textbf{23} (2010), no. 2, 345–382.

    \bibitem[KRS15]{KRS15}
    Klainerman, Sergiu; Rodnianski, Igor; Szeftel, Jeremie:
    The bounded $L^2$ curvature conjecture.
    \textit{Invent. Math.} \textbf{202} (2015), no. 1, 91–217.

    \bibitem[KOSS22]{KOSS22}
    Kunzinger, Michael; Ohanyan, Argam; Schinnerl, Benedict; Steinbauer, Roland:
    The Hawking-Penrose singularity theorem for $\mathrm{C}^1$-Lorentzian metrics.
    \textit{Comm. Math. Phys.} \textbf{391} (2022), no. 3, 1143–1179.

    \bibitem[KSV15]{KSV15}
    Kunzinger, Michael; Steinbauer, Roland; Vickers, James A.:
    The Penrose singularity theorem in regularity $\mathrm{C}^{1,1}$.
    \textit{Classical Quantum Gravity} \textbf{32} (2015), no. 15, 155010, 12 pp.

    \bibitem[KSSV15]{KSSV15}
    Kunzinger, Michael; Steinbauer, Roland; Stojkovi\'c, Milena; Vickers, James A.:
    Hawking's singularity theorem for $\mathrm{C}^{1,1}$-metrics.
    \textit{Classical Quantum Gravity} \textbf{32} (2015), no. 7, 075012, 19 pp.

    \bibitem[LLS21]{LLS21}
    Lange, Christian; Lytchak, Alexander; S\"amann, Clemens:
    Lorentz meets Lipschitz.
    \textit{Adv. Theor. Math. Phys.} \textbf{25} (2021), no. 8, 2141–2170.

    \bibitem[Le23]{Le23}
    Le, Pengyu:
    Lorentz polarisation and isoperimetric inequality in Minkowski spacetime.
    arXiv:2307.03301 [math.DG] (2023).

    \bibitem[Li25]{Li25}
    Li, Junbin:
    Interior instability of naked singularities of a scalar field.
    arXiv:2508.07655 [gr-qc] (2025).

    \bibitem[LL18]{LL18}
    Li, Junbin; Liu, Jue:
    A robust proof of the instability of naked singularities of a scalar field in spherical symmetry.
    \textit{Comm. Math. Phys.} \textbf{363} (2018), no. 2, 561–578.

    \bibitem[LL22]{LL22}
    Li, Junbin; Liu, Jue:
    Instability of spherical naked singularities of a scalar field under gravitational perturbations.
    \textit{J. Differential Geom.} \textbf{120} (2022), no. 1, 97–198.

    \bibitem[Lin20]{Lin20}
    Ling, Eric:
    The big bang is a coordinate singularity for $k=-1$ inflationary FLRW spacetimes.
    \textit{Found. Phys.} \textbf{50} (2020), no. 5, 385–428.

    \bibitem[Lin26]{Lin26}
    Ling, Eric:
    The $\mathrm{C}^0$-inextendibility of some spatially flat FLRW spacetimes.
    \textit{Gen. Relativ. Gravit.} \textbf{58} (2026), no. 5, Paper No. 52, 11 pp.

    \bibitem[Lu18]{Lu18}
    Luk, Jonathan:
    Weak null singularities in general relativity.
    \textit{J. Amer. Math. Soc.} \textbf{31} (2018), no. 1, 1–63.

    \bibitem[LO19a]{LO19a}
    Luk, Jonathan; Oh, Sung-Jin:
    Strong cosmic censorship in spherical symmetry for two-ended asymptotically flat initial data I. The interior of the black hole region.
    \textit{Ann. of Math.} (2) \textbf{190} (2019), no. 1, 1–111.

    \bibitem[LO19b]{LO19b}
    Luk, Jonathan; Oh, Sung-Jin:
    Strong cosmic censorship in spherical symmetry for two-ended asymptotically flat initial data II: the exterior of the black hole region.
    \textit{Ann. PDE} \textbf{5} (2019), no. 1, Paper No. 6, 194 pp.

    \bibitem[LS26]{LS26}
    Luk, Jonathan; Sbierski, Jan:
    The formation of a weak null singularity in the interior of generic rotating black holes.
    arXiv:2604.04877 [gr-qc] (2026).

    \bibitem[Mie24]{Mie24}
    Miethke, Benedikt:
    $\mathrm{C}^0$-inextendibility of the Kasner spacetime.
    arXiv:2408.05257 [gr-qc] (2024).

    \bibitem[Min19]{Min19}
    Minguzzi, Ettore:
    Causality theory for closed cone structures with applications.
    \textit{Rev. Math. Phys.} \textbf{31} (2019), no. 5, 1930001, 139 pp.

    \bibitem[MS08]{MS08}
    Minguzzi, Ettore; S\'anchez, Miguel:
    The causal hierarchy of spacetimes.
    \textit{Recent developments in pseudo-Riemannian geometry}, 299–358.
    ESI Lect. Math. Phys.
    European Mathematical Society (EMS), Zürich, 2008

    \bibitem[MS19]{MS19}
    Minguzzi, E.; Suhr, S.:
    Some regularity results for Lorentz-Finsler spaces.
    Ann. Global Anal. Geom. 56 (2019), no. 3, 597–611.

    \bibitem[My78]{My78}
    Myrheim Jan:
    Statistical geometry.
    CERN preprint Report No. CERN-TH-2538, 1978, 13pp.

    \bibitem[P79]{P79}
    Penrose, Roger:
    Singularities and time-asymmetry.
    \textit{General relativity.}
    An Einstein centenary survey. Edited by S. W. Hawking and W. Israel. Cambridge University Press, Cambridge-New York, 1979, 581–638.

    \bibitem[S\"a16]{Sa16}
    S\"amann, Clemens:
    Global hyperbolicity for spacetimes with continuous metrics.
    \textit{Ann. Henri Poincar\'e} \textbf{17} (2016), no. 6, 1429–1456.

    \bibitem[Sb18a]{Sb18a}
    Sbierski, Jan:
    The $\mathrm{C}^0$-inextendibility of the Schwarzschild spacetime and the spacelike diameter in Lorentzian geometry.
    \textit{J. Differential Geom.} \textbf{108} (2018), no. 2, 319–378.

    \bibitem[Sb18b]{Sb18b}
    Sbierski, Jan:
    On the proof of the $\mathrm{C}^0$-inextendibility of the Schwarzschild spacetime.
    \textit{J. Phys. Conf. Ser.} \textbf{968} (2018), 012012, 16 pp.

    \bibitem[Sb22]{Sb22}
    Sbierski, Jan:
    On holonomy singularities in general relativity and the $\mathrm{C}^{0,1}_{\text{loc}}$-inextendibility of space-times.
    \textit{Duke Math. J.} \textbf{171} (2022), no. 14, 2881–2942.

    \bibitem[Sb23]{Sb23}
    Sbierski, Jan:
    The $\mathrm{C}^0$-inextendibility of a class of FLRW spacetimes.
    arXiv:2312.07443 [gr-qc] (2023).

    \bibitem[Sb25]{Sb25}
    Sbierski, Jan:
    Lipschitz inextendibility of weak null singularities from curvature blow-up.
    Invent. math. (2025).

    \bibitem[So25]{So25}
    Song, Yuefeng:
    Weak null singularity for the Einstein-Euler system.
    arXiv:2506.16635 [gr-qc] (2025).

    \bibitem[Wj19]{Wj19}
    Wang, Jinzhao:
    Geometry of small causal diamonds.
    \textit{Phys. Rev. D} \textbf{100} (2019), no. 6, 064020, 14 pp.

    \bibitem[Wq12]{Wq12}
    Wang, Qian:
    Improved breakdown criterion for Einstein vacuum equations in CMC gauge.
    \textit{Comm. Pure Appl. Math.} \textbf{65} (2012), no. 1, 21–77.



\end{thebibliography}
\end{document}